\pdfoutput=1
\documentclass[iop, apj, twocolappendix, numberedappendix]{emulateapj}

\newif\iflocal
\localfalse

\usepackage{xspace}
\usepackage{amsmath}
\usepackage{framed} 
\usepackage{txfonts}
\usepackage{epstopdf}
\usepackage{color}
\usepackage{rotating}
\usepackage{natbib}
\usepackage{ulem}
\usepackage{xspace}
\usepackage[colorlinks=true,urlcolor=black,linkcolor=blue,citecolor=blue]{hyperref}
\usepackage{sistyle}
\SIthousandsep{,}

\iflocal

\newcommand{\kpch}{\>{h^{-1}{\rm kpc}}}
\newcommand{\mpch}{\>h^{-1}{\rm {Mpc}}}

\newcommand{\msunh}{\>h^{-1} M_\odot}

\def\gcm3{\mathrm{g} / \mathrm{cm}^3}

\def\LCDM{$\Lambda$CDM\ }

\def\mvir{M_{\rm vir}}
\def\rvir{R_{\rm vir}}

\def\mtom{M_{\rm 200m}}
\def\rtom{R_{\rm 200m}}

\def\vtom{v_{\rm 200m}}
\def\ntom{N_{\rm 200m}}
\def\nutom{\nu_{\rm 200m}}

\def\rtoc{R_{\rm 200c}}

\def\mdelta{M_{\Delta}}
\def\rdelta{R_{\Delta}}

\def\rhoc{\rho_{\rm c}}
\def\rhom{\rho_{\rm m}}

\def\sparta{\textsc{Sparta}\xspace}
\def\shellfish{\textsc{Shellfish}\xspace}
\def\gotetra{\textsc{gotetra}\xspace}
\def\colossus{\textsc{Colossus}\xspace}

\def\gtsima{$\; \buildrel > \over \sim \;$}
\def\ltsima{$\; \buildrel < \over \sim \;$}
\def\prosima{$\; \buildrel \propto \over \sim \;$}
\def\gsim{\lower.7ex\hbox{\gtsima}}
\def\lsim{\lower.7ex\hbox{\ltsima}}
\def\simgt{\lower.7ex\hbox{\gtsima}}
\def\simlt{\lower.7ex\hbox{\ltsima}}
\def\simpr{\lower.7ex\hbox{\prosima}}

\def\tdyn{t_{\rm dyn}}

\def\gammadk{\Gamma_{\rm DK14}}
\def\gammadyn{\Gamma_{\rm dyn}}

\def\rsp{R_{\rm sp}}
\def\rrsp{r_{\rm sp}}
\def\tsp{t_{\rm sp}}
\def\rspmean{R_{\rm sp}^{\rm mn}}
\def\rspmed{R_{\rm sp}^{50\%}}
\def\rspsf{R_{\rm sp}^{75\%}}
\def\rspes{R_{\rm sp}^{87\%}}

\def\rsptom{\rsp/\rtom}

\def\msp{M_{\rm sp}}
\def\mmsp{m_{\rm sp}}
\def\mspmean{M_{\rm sp}^{\rm mn}}
\def\mspmed{M_{\rm sp}^{50\%}}

\def\msptom{\msp/\mtom}

\def\deltasp{\Delta_{\rm sp}}

\def\deltaspes{\Delta_{\rm sp}^{87\%}}

\def\xsp{X_{\rm sp}}

\def\gammarsp{$\Gamma$--$\rsp$ relation\xspace}
\def\gammarsps{$\Gamma$--$\rsp$ relations\xspace}
\def\nursp{$\nu$--$\rsp$ relation\xspace}
\def\nugamma{$\nu$--$\Gamma$ relation\xspace}

\usepackage{etoolbox}
\makeatletter
\patchcmd{\NAT@citex}
  {\@citea\NAT@hyper@{\NAT@nmfmt{\NAT@nm}\NAT@date}}
  {\@citea\NAT@nmfmt{\NAT@nm}\NAT@hyper@{\NAT@date}}
  {}
  {}
\patchcmd{\NAT@citex}
  {\@citea\NAT@hyper@{%
     \NAT@nmfmt{\NAT@nm}%
     \hyper@natlinkbreak{\NAT@aysep\NAT@spacechar}{\@citeb\@extra@b@citeb}%
     \NAT@date}}
  {\@citea\NAT@nmfmt{\NAT@nm}%
   \NAT@aysep\NAT@spacechar%
   \NAT@hyper@{\NAT@date}}
  {}
  {}
\patchcmd{\NAT@citex}
  {\@citea\NAT@hyper@{%
     \NAT@nmfmt{\NAT@nm}%
     \hyper@natlinkbreak{\NAT@spacechar\NAT@@open\if*#1*\else#1\NAT@spacechar\fi}%
       {\@citeb\@extra@b@citeb}%
     \NAT@date}}
  {\@citea\NAT@nmfmt{\NAT@nm}%
   \NAT@spacechar\NAT@@open\if*#1*\else#1\NAT@spacechar\fi%
   \NAT@hyper@{\NAT@date}}
  {}
  {}
\makeatother

\else

\fi

\shorttitle{The Splashback Radius}
\shortauthors{Diemer et al.}

\journalinfo{The Astrophysical Journal, {\rm 843:140 (16pp), 2014 July 10}}
\submitted{Received 2017 March 28; revised 2017 June 2; accepted 2017 June 11; published 2017 July 14}

\begin{document}


\iflocal
\def\figdir{figs}
\else
\def\figdir{.}
\fi


\defcitealias{diemer_13_scalingrel}{DKM13}
\defcitealias{diemer_14_profiles}{DK14}
\defcitealias{diemer_15_cm}{DK15}
\defcitealias{diemer_17_sparta}{Paper I}


\title{The splashback radius of halos from particle dynamics. II. Dependence on mass,\\ accretion rate, redshift, and cosmology}
\author{Benedikt Diemer\altaffilmark{1}, Philip Mansfield\altaffilmark{2,3}, Andrey V. Kravtsov\altaffilmark{2,3}, Surhud More\altaffilmark{4}}

\affil{
$^1$ Institute for Theory and Computation, Harvard-Smithsonian Center for Astrophysics, 60 Garden St., Cambridge, MA 02138, USA; \href{mailto:benedikt.diemer@cfa.harvard.edu}{benedikt.diemer@cfa.harvard.edu}\\
$^2$ Department of Astronomy and Astrophysics, The University of Chicago, Chicago IL 60637, USA \\
$^3$ Kavli Institute for Cosmological Physics, The University of Chicago, Chicago IL 60637, USA \\
$^4$ Kavli Institute for the Physics and Mathematics of the Universe (WPI), The University of Tokyo,\\ 5-1-5 Kashiwanoha, Kashiwa-shi, Chiba, 277-8583, Japan
}


\begin{abstract}
The splashback radius $\rsp$, the apocentric radius of particles on their first orbit after falling into a dark matter halo, has recently been suggested as a physically motivated halo boundary that separates accreting from orbiting material. Using the \sparta code presented in Paper I, we analyze the orbits of billions of particles in cosmological simulations of structure formation and measure $\rsp$ for a large sample of halos that span a mass range from dwarf galaxy to massive cluster halos, reach redshift $8$, and include WMAP, Planck, and self-similar cosmologies. We analyze the dependence of $\rsptom$ and $\msptom$ on the mass accretion rate $\Gamma$, halo mass, redshift, and cosmology. The scatter in these relations varies between $0.02$ and $0.1$ dex. While we confirm the known trend that $\rsptom$ decreases with $\Gamma$, the relationships turn out to be more complex than previously thought, demonstrating that $\rsp$ is an independent definition of the halo boundary that cannot trivially be reconstructed from spherical overdensity definitions. We present fitting functions for $\rsptom$ and $\msptom$ as a function of accretion rate, peak height, and redshift, achieving an accuracy of 5\% or better everywhere in the parameter space explored. We discuss the physical meaning of the distribution of particle apocenters and show that the previously proposed definition of $\rsp$ as the radius of the steepest logarithmic density slope encloses roughly three-quarters of the apocenters. Finally, we conclude that no analytical model presented thus far can fully explain our results.
\end{abstract}



\section{Introduction}
\label{sec:intro}

According to our current understanding of structure formation, cold dark matter hierarchically collapses into condensations called halos. Baryons follow this collapse on large scales and cool at the centers of halos to form galaxies \citep{rees_77, silk_77, white_78}. There is a tight connection between the masses and evolutionary histories of galaxies and their halos, as demonstrated by the success of various classes of models for the galaxy-halo connection that have been put forward over the past decades. For example, subhalo abundance matching \citep{kravtsov_04, tasitsiomi_04_sham, vale_04, conroy_06, conroy_09_shmr, moster_10_shmr, behroozi_13_shmr} assigns galaxies to halos based on rank orderings of stellar mass and some halo property, such as mass. Halo occupation distributions \citep{peacock_00, seljak_00_wl, berlind_02, cooray_02} assign one or multiple galaxies to a halo based on its mass or other properties \citep{hearin_16_decoratedhod}. Finally, semi-analytical models \citep{kauffmann_93, somerville_99, bower_06, guo_10} attempt to describe the sophisticated mechanisms of galaxy formation but are ultimately based on merger trees that represent the evolutionary histories of halos. Even in hydrodynamic simulations designed to follow the formation of galaxies from first principles, certain parameters are sometimes explicitly tied to the halo mass or radius --- for example the seeding of black holes \citep{vogelsberger_13_model} or stellar wind velocities \citep{dave_16_mufasa}.

Thus, the models described above share one important caveat: they depend upon a particular definition of the halo boundary and mass. The most widely accepted definition is for the halo radius to enclose some overdensity $\Delta$ such that
\begin{equation}
\label{eq:so}
\mdelta = \frac{4\pi}{3} \Delta \rho_{\rm ref} \rdelta^3
\end{equation}
where $\rho_{\rm ref}$ is either the critical or mean matter density of the universe \citep[e.g.][]{cole_96_halostructure}. This spherical overdensity definition has a number of manifest advantages: radius and mass are trivially related via the reference density, $\mdelta$ can be measured in both simulations and observations by counting the mass included in shells of increasing radius, and re-scaling halo radii by $\rdelta$ leads to a self-similar form of the density profile that can approximately be described as a function of only mass and a concentration parameter \citep{navarro_95, navarro_96, navarro_97_nfw, navarro_04, burkert_95, cole_96_halostructure}. 

By contrast, spherical overdensity radii and masses suffer from a number of issues. First, the extent to which the profiles are self-similar at different masses and redshifts depends on the somewhat arbitrarily chosen overdensity threshold \citep{diemer_14_profiles, diemer_15_cm}. One can derive a so-called virial overdensity of $\Delta_{\rm vir} = 178$ from arguments based on the collapse of an isolated top-hat overdensity in an $\Omega_{\rm m} = 1$ universe \citep{gunn_72_sphericalcollapse, peebles_80_structurebook, lacey_93_collapse}, where the overdensity evolves with time in $\Lambda$CDM cosmologies \citep[e.g.,][]{lahav_91_lambda_clusters}. However, the peaks in the initial Gaussian random field are not in a top-hat shape \citep{dalal_08, dalal_10}, the particles do not instantaneously virialize as assumed in the model \citep[e.g.][]{shaw_06, sanchezconde_07, ludlow_12}, and halos do not form in isolation, creating complicated density fields that extend well past $\rvir$ \citep{prada_06_outerregions, hayashi_08, oguri_11, diemer_14_profiles}. One manifestation of this extended structure is that subhalos falling into a more massive host begin to lose mass long before they cross the host's $\rvir$ \citep{behroozi_14_infall, penarrubia_17}. Finally, spherical overdensity masses can grow unphysically despite a constant halo density profile because the reference density decreases with cosmic time, an effect called pseudo-evolution \citep{diemand_05, cuesta_08_infall, diemer_13_pe, zemp_14, more_15}.

In order to mitigate these issues, a number of alternative mass definitions have been put forward. The most popular of these is the friends-of-friends (FOF) mass \citep{davis_85_clustering, jenkins_01_mfunc}. Although appealingly simple, this algorithm relies on a somewhat arbitrarily chosen linking length parameter, and, for common choices of this parameter, FOF groups can include neighboring halos \citep{white_01_mass_definitions}. Furthermore, FOF masses have been shown to suffer from dependencies on mass resolution and halo concentration \citep{more_11_fof, benson_17_unprocessed}. Another alternative was suggested by \citet{cuesta_08_infall}, who argued for the radius where the average radial velocity changes from outflowing to infalling. This radius, however, is not clearly defined in some low-mass halos and encloses a large amount of matter falling toward the halo for the first time that arguably should not be included \citep{diemer_14_profiles}. \citet{anderhalden_11_totalmass} suggested counting all particles that ever entered the halo, a definition that suffers from similar theoretical issues. The ORIGAMI algorithm \citep{falck_12, neyrinck_12} defines halos by identifying particles that have switched positions with other particles along three orthogonal axes. Theoretical considerations aside, the most important issue with all of these mass definitions is that they cannot be measured in the real universe.

Recently, it has been argued that a more natural halo boundary is provided by the splashback radius, $\rsp$, the radius where particles reach the apocenter of their first orbit after infall \citep{diemer_14_profiles, adhikari_14, more_15, mansfield_17}. The theoretical inspiration for this definition is provided by the spherical collapse model, where spherically symmetric shells of matter successively fall onto an initial power-law density perturbation, creating a power-law inner density profile \citep{fillmore_84, bertschinger_85, mohayaee_06, ascasibar_07, diemand_08, vogelsberger_11, lithwick_11, adhikari_14, shi_16}. Particles at the apocenter of their first orbit pile up due to their low radial velocity, creating a caustic that manifests itself as a sharp drop in the density profile. This so-called splashback radius represents a clear boundary between matter orbiting in the halo and matter on a first infall toward the halo. 

\begin{deluxetable}{ll}
\tablecaption{Definitions of the Symbols Used in This Paper
\label{table:defs}}
\tablewidth{0pt}
\tablehead{
\colhead{Symbol} &
\colhead{Meaning}
}
\startdata
$\rhom$ & Mean matter density of the universe \\
$\rhoc$ & Critical density of the universe \\
$\Omega_{\rm m}$ & Fractional matter density, $\Omega_{\rm m} \equiv \rhom / \rhoc$ \\
$r$ & Some radius in physical units, measured from the halo center \\
$R$ & A particular definition of the halo boundary \\
$\Delta$ & An overdensity with respect to either $\rhom$ or $\rhoc$ \\
$\rtom$ & Radius enclosing an overdensity of $200 \times \rhom$ \\
$\rtoc$ & Radius enclosing an overdensity of $200 \times \rhoc$ \\
$\rvir$ & $\rdelta$ with varying overdensity \citep{bryan_98_virial} \\
$\mtom$ & Mass inside $\rtom$ \\
$N_{\Delta}$ & Number of particles inside $\rdelta$, e.g. $N_{\rm 200m}$ \\
$\nu$ &  Peak height, $\nu \equiv \nutom = \delta_{\rm c} / \sigma(\mtom, z)$ \\
$r_{\rm s}$ & Scale radius of an NFW profile \\
$c_{\Delta}$ & Concentration, $c_{\Delta} \equiv \rdelta / r_{\rm s}$ \\
$\rrsp$ & Splashback radius of a particle \\
$\mmsp$ & Splashback mass of a particle, i.e. $M(<\rrsp)$ \\
$\rsp$ & Splashback radius of a halo \\
$\msp$ & Splashback mass of a halo \\
$\deltasp$ & Splashback overdensity wrt. $\rhom$, $\deltasp \equiv 3 \msp / (4 \pi \rsp^3) / \rho_{\rm m}$ \\
$\rspmean$ & $\rsp$ defined as the mean of the particle $\rrsp$ \\
$\rspmed$ & $\rsp$ defined as the median of the particle  $\rrsp$ \\
$\rspsf$ & $\rsp$ defined as the 75th percentile of the particle  $\rrsp$ \\
$\rsp^*$ & Summary symbol for multiple definitions of $\rsp$ \\
$v_{\Delta}$ & Circular velocity, $v_{\Delta} \equiv \sqrt{G\mdelta / \rdelta}$ \\
$t_{\rm dyn}$ & Dynamical time or crossing time, $t_{\rm dyn} \equiv 2 \rtom / \vtom$ \\
$s$ & Instantaneous mass accretion rate, $d \log(M) / d \log(a)$ \\
$\gammadyn$ & Mass accretion rate  over one $t_{\rm dyn}$, $\Delta \log(M) / \Delta \log(a)$ \\
$\gammadk$ & Mass accretion rate as defined in \citet{diemer_14_profiles}
\enddata
\end{deluxetable}

The sharp drop in stacked halo density profiles at the splashback radius was recently detected in cosmological simulations \citep{diemer_14_profiles}, and its location was shown to primarily depend on mass accretion rate. \citet{adhikari_14} reproduced this dependence with a simple theoretical model, making a convincing case for the connection between the splashback radius and the density drop \citep[see also][]{shi_16}. \citet{more_15} adopted the definition of $\rsp$ as the radius where the density profile reaches its steepest slope and investigated its dependence on mass accretion rate and redshift. Finally, \citet[][see also \citealt{adhikari_16_df} and \citealt{baxter_17}]{more_16} detected a sharp drop in the stacked density profiles of galaxy cluster members at the splashback radius, though such observations are complicated by the systematics of the cluster identification method \citep[][see also \citealt{rines_13}, \citealt{tully_15}, \citealt{patej_16}, and \citealt{umetsu_17} for hints of the splashback radius in observations of individual clusters and weak lensing signals]{zu_17, busch_17}.

\begin{deluxetable*}{lccccccccccl}
\tablecaption{$N$-Body Simulations
\label{table:sims}}
\tablewidth{0pt}
\tablehead{
\colhead{Name} &
\colhead{$L$} &
\colhead{$N^3$} &
\colhead{$m_{\rm p}$} &
\colhead{$\epsilon$} &
\colhead{$\epsilon / (L / N)$} &
\colhead{$z_{\rm initial}$} &
\colhead{$z_{\rm final}$} &
\colhead{$N_{\rm snaps}$} &
\colhead{$z_{\rm f-snap}$} &
\colhead{Cosmology} &
\colhead{Reference}
}
\startdata
L2000        & $2000$  & $1024^3$ & $5.6 \times 10^{11}$  & $65$   & $1/30$  & $49$  & $0$ & $100$ & $20$ & $WMAP$ (Bolshoi) & \citetalias{diemer_15_cm} \\
L1000        & $1000$  & $1024^3$ & $7.0 \times 10^{10}$  & $33$   & $1/30$  & $49$  & $0$ & $100$ & $20$ & $WMAP$ (Bolshoi) & \citetalias{diemer_13_scalingrel} \\
L0500        & $500$   & $1024^3$ & $8.7 \times 10^{9}$   & $14$   & $1/35$  & $49$  & $0$ & $100$ & $20$ & $WMAP$ (Bolshoi) & \citetalias{diemer_14_profiles} \\
L0250        & $250$   & $1024^3$ & $1.1 \times 10^{9}$   & $5.8$  & $1/42$  & $49$  & $0$ & $100$ & $20$ & $WMAP$ (Bolshoi) & \citetalias{diemer_14_profiles} \\
L0125        & $125$   & $1024^3$ & $1.4 \times 10^{8}$   & $2.4$  & $1/51$  & $49$  & $0$ & $100$ & $20$ & $WMAP$ (Bolshoi) & \citetalias{diemer_14_profiles} \\
L0063        & $62.5$  & $1024^3$ & $1.7 \times 10^{7}$   & $1.0$  & $1/60$  & $49$  & $0$ & $100$ & $20$ & $WMAP$ (Bolshoi) & \citetalias{diemer_14_profiles} \\
L0031        & $31.25$ & $1024^3$ & $2.1 \times 10^{6}$   & $0.25$ & $1/122$ & $49$  & $2$ & $64$  & $20$ & $WMAP$ (Bolshoi) & \citetalias{diemer_15_cm} \\
L0500-Planck & $500$   & $1024^3$ & $1.0 \times 10^{10}$  & $14$   & $1/35$  & $49$  & $0$ & $100$ & $20$ & $Planck$ & \citetalias{diemer_15_cm} \\
L0250-Planck & $250$   & $1024^3$ & $1.3 \times 10^{9}$   & $5.8$  & $1/42$  & $49$  & $0$ & $100$ & $20$ & $Planck$ & \citetalias{diemer_15_cm} \\
L0125-Planck & $125$   & $1024^3$ & $1.6 \times 10^{8}$   & $2.4$  & $1/51$  & $49$  & $0$ & $100$ & $20$ & $Planck$ & \citetalias{diemer_15_cm} \\
L0100-PL-1.0 & $100$   & $1024^3$ & $2.6 \times 10^{8}$   & $0.5$  & $1/195$ & $119$ & $2$ & $64$  & $20$ & Self-similar, $n=-1.0$ & \citetalias{diemer_15_cm} \\
L0100-PL-2.5 & $100$   & $1024^3$ & $2.6 \times 10^{8}$   & $1.0$  & $1/98$  & $49$  & $0$ & $100$  & $20$ & Self-similar, $n=-2.5$ & \citetalias{diemer_15_cm}
\enddata
\tablecomments{The $N$-body simulations used in this paper. Here $L$ denotes the box size in comoving $\mpch$, $N^3$ is the number of particles, $m_{\rm p}$ the particle mass in $\msunh$, $\epsilon$ is the force-softening length in physical $\kpch$, $z_{\rm initial}$ and $z_{\rm final}$ are the redshift range of the simulation, $N_{\rm snaps}$ is the number of snapshots written to disk, and $z_{\rm f-snap}$ is the redshift of the first snapshot. The references correspond to \citet[][\citetalias{diemer_13_scalingrel}]{diemer_13_scalingrel}, \citet[][\citetalias{diemer_14_profiles}]{diemer_14_profiles}, and \citet[][\citetalias{diemer_15_cm}]{diemer_15_cm}. More details on our logic for choosing force resolutions are given in \citetalias{diemer_14_profiles}.}
\end{deluxetable*}

\begin{deluxetable*}{lccccccccll}
\tablecaption{Cosmological Parameters
\label{table:cosmo}}
\tablewidth{0pt}
\tablehead{
\colhead{Cosmology} &
\colhead{$H_0$} &
\colhead{$\Omega_{\rm m}$} &
\colhead{$\Omega_{\rm \Lambda}$} &
\colhead{$\Omega_{\rm b}$} &
\colhead{$\Omega_{\rm k}$} &
\colhead{$\Omega_{\rm \nu}$} &
\colhead{$\sigma_8$} &
\colhead{$n_{\rm s}$} &
\colhead{$P(k)$} &
\colhead{Reference}
}
\startdata
$WMAP$ (Bolshoi)                        & $70$ & $0.27$ & $0.73$ & $0.0469$ & $0$ & $0$ & $0.82$  & $0.95$   & CAMB &  \citet{klypin_11_bolshoi}, \citet{komatsu_11} \\
$Planck$                         & $67$ & $0.32$ & $0.68$ & $0.0491$ & $0$ & $0$ & $0.834$ & $0.9624$ & CAMB & \citet{planck_14_cosmology} \\
Self-similar                   & $70$ & $1$    & $0$    & $0$      & $0$ & $0$ & $0.82$  & ...       & $P(k) \propto k^n$ & ...
\enddata
\tablecomments{Cosmological parameters of the $N$-body simulations listed in Table \ref{table:sims}. The Bolshoi cosmology roughly corresponds to the $WMAP7$ cosmology of \citet{komatsu_11}. The $Planck$ values correspond to the $Planck$-only best-fit values given in Table 2 of \citet{planck_14_cosmology}. Some of the parameters in both the $Planck$ and Bolshoi cosmologies are rounded for convenience. The initial matter power spectrum for the Bolshoi and $Planck$ cosmologies was computed using the Boltzmann code \textsc{Camb} \citep{lewis_00_camb}.}
\end{deluxetable*}

All of the theoretical and observational work discussed above has been based on the definition of $\rsp$ as the radius where the logarithmic slope of the density profile is steepest. While this definition is intuitive, the radius of the steepest slope is affected by a trade-off between the sharply falling inner profile and the outer infall region. Thus, it is not clear what fraction of particles actually reach their orbital apocenter inside the radius of the steepest slope and whether this fraction is universal across halo masses, redshifts, and cosmologies. Moreover, \citet{mansfield_17} showed that substructure can wipe out the signature of splashback in simulated density profiles and leads to a significant bias in the splashback radius measured from stacked density profiles. Finally, the scatter in the $\rsp$ distribution cannot be determined from stacked density profiles.

For all of these reasons, it is desirable to measure $\rsp$ in individual simulated halos using a method that does not rely on spherically averaged density profiles. \citet{mansfield_17} performed such measurements using the full three-dimensional density information to obtain nonspherical splashback shells. While their method relies only on the density field at a given time, it demands relatively well-resolved halos with more than \num{50000} particles and can fail for the slowest accreting fraction of halos \citep{mansfield_17}. In order to measure $\rsp$ in less well-resolved systems, \citet[][hereafter \citetalias{diemer_17_sparta}]{diemer_17_sparta} suggested an algorithm based on the apocenter passages of individual particles. This method necessarily uses all of the snapshots of a simulation but was shown to converge for halos resolved by as few as $1000$ particles. 

In this second paper in the series, we investigate the relation between the $\rsp$ and $\msp$ of individual halos and their spherical overdensity mass, accretion rate, redshift, and cosmology. In order to facilitate the use of $\rsp$ as a practical definition of the halo boundary, we provide accurate fitting functions for these dependencies. While the particle apocenters are not directly observable, we discuss the connection of our new $\rsp$ measurements to results based on stacked density profiles. 

The paper is organized as follows. In Section~\ref{sec:methods}, we briefly summarize our simulations and algorithm, referring the reader to \citetalias{diemer_17_sparta} for details. We show our results in Section~\ref{sec:results}, and compare them to previous work and theoretical models in Section~\ref{sec:comp}. We further discuss the implications of our results in Section~\ref{sec:discussion} and summarize our conclusions in Section~\ref{sec:conclusion}.

Throughout the paper, we adopt the same symbols as in \citetalias{diemer_17_sparta} which are summarized in Table~\ref{table:defs}. Any input quantities to the fitting functions for $\rsp$, such as peak height and mass accretion rate, are defined in terms of conventional masses and radii that can be measured by a standard halo finder. While theoretical models typically refer to the instantaneous accretion rate $s \equiv d \log(M) / d \log(a)$, this quantity cannot be measured in simulation data due to the noisy nature of mass accretion histories. Thus, \citet{diemer_14_profiles} defined the mass accretion rate over a finite range of time,
\begin{equation}
\label{eq:gammadk14}
\Gamma(a_1) \equiv \frac{\Delta \log(M)}{\Delta \log(a)} = \frac{\log(M_1) - \log(M_0)}{\log(a_1) - \log(a_0)} \,,
\end{equation}
where $M = \mvir$ and the $a_0$-$a_1$ pairs were chosen manually to correspond to roughly a crossing time \citep[see also][]{lau_15, more_15, mansfield_17}. As in \citetalias{diemer_17_sparta}, we instead choose $M = \mtom$ and measure the accretion rate over one dynamical time, $a_1 \equiv a(t - \tdyn)$. The dynamical time used in this definition depends only on the chosen overdensity and cosmology, not on the properties of individual halos \citepalias{diemer_17_sparta}. We emphasize that one has to be careful when interpreting mass accretion rates in terms of the growth of the physical density profile. Pseudo-evolution, i.e. spurious growth due to the changing definition of the overdensity with redshift, contributes to changes in $\mtom$. For example, for a static, non-evolving Navarro-Frenk-White (NFW) profile, the ``accretion rate'' is $\Gamma \approx 0.5$ (regardless of redshift or halo mass). Thus, halos with $\Gamma < 0.5$ are maintaining the same physical mass profile within $\rtom$, or are even losing mass (e.g. due to tidal disruption as they approach another halo).

 
\section{Methods}
\label{sec:methods}

\begin{figure*}
\centering
\includegraphics[trim = 0mm 0mm 0mm 0mm, clip, width=18cm]{\figdir/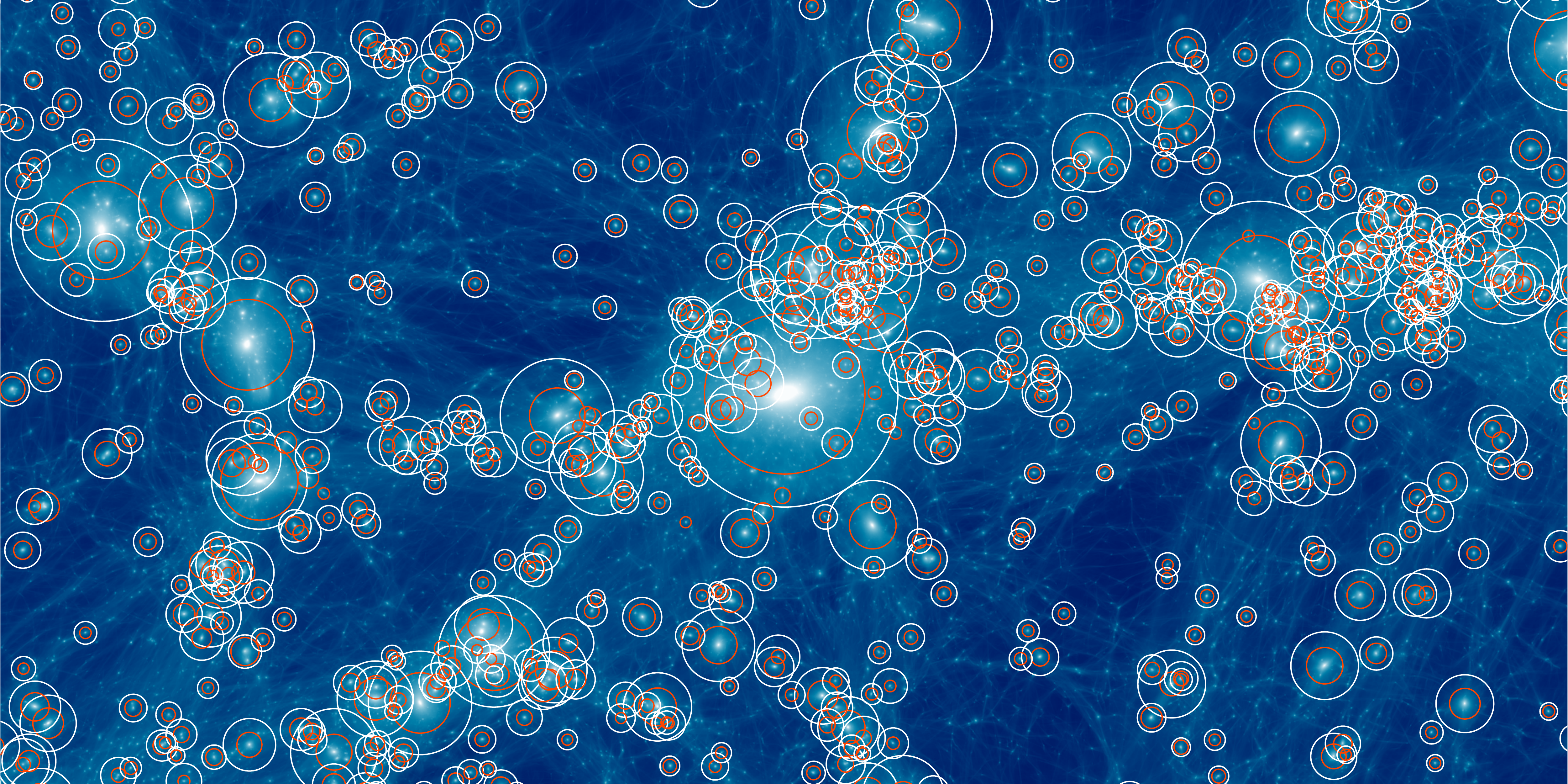}
\caption{Comparison of conventional ``virial'' and splashback radii ($\rvir$ and $\rsp$, shown as orange and white circles). The image shows the projected density through a slice in the L0125 simulation that is $30 \mpch$ wide and deep and $15 \mpch$ tall. The density field is visualized using the \gotetra code (Mansfield et al. 2017, in preparation). Radii are shown for all halos with $\ntom \geq 1000$ (equivalent to a mass of $1.4 \times 10^{11} \msunh$), and the mass of the central halo is $\mtom = 1.2 \times 10^{14} \msunh$ (corresponding to almost a million particles). The splashback radii shown are defined as $\rspes$ which corresponds most closely to the density drop measured by the \shellfish code (see Section~\ref{sec:comp:shellfish}). For a small fraction of halos, \sparta could not determine a splashback radius because they had recently been subhalos.}
\label{fig:viz}
\end{figure*}

In this section, we describe the $N$-body simulations used for this project and give a brief summary of the \sparta algorithm, referring the reader to \citetalias{diemer_17_sparta} for details.

\subsection{$N$-Body Simulations}
\label{sec:method:sims}

Our results are based on a suite of dissipationless \LCDM simulations of different box sizes and cosmologies (Table~\ref{table:sims}). Our fiducial cosmology is the same as that of the Bolshoi simulation \citep{klypin_11_bolshoi}, but we also use simulations of the $Planck$ cosmology in order to investigate the cosmology dependence of the splashback radius (see Table~\ref{table:cosmo}). The initial conditions for the simulations were generated using the second-order Lagrangian perturbation theory code \textsc{2LPTic} \citep{crocce_06_2lptic}. The simulations were started at redshift $z = 49$, sufficiently high to avoid transient effects \citep{crocce_06_2lptic}, and were run with the publicly available code \textsc{Gadget2} \citep{springel_05_gadget2}. In \citetalias{diemer_17_sparta}, we showed that the number of saved snapshots (generally $100$) is sufficient for the \sparta algorithm to give reliable results.

We used the phase-space halo finder \textsc{Rockstar} \citep{behroozi_13_rockstar} to extract halos and subhalos from the snapshots of each simulation and the \textsc{Consistent-Trees} code \citep{behroozi_13_trees} to establish subhalo relations and assemble merger trees. We note that the halo catalogs and merger trees used in this paper are based on $\rtom$ as the halo radius. This definition matters because halos whose centers lie inside $\rtom$ of another, larger halo are considered subhalos and are treated rather differently (Section~\ref{sec:method:sparta}). \textsc{Rockstar} computes $\rtom$ using only bound particles in order to avoid spurious contributions from their hosts. While the merger trees are based on these bound-only radii, we generally use $\rtom$ as computed from all particles, bound and unbound, and explicitly state when we are using bound-only masses and radii. For the vast majority of host halos, the difference between the two masses is small.

\subsection{The \sparta algorithm}
\label{sec:method:sparta}

In each host halo, we track all particles as they fall into the halo for the first time and record whether a particle was part of a subhalo at infall. Thereafter, we follow the particle's trajectory and, at the apocenter of its first orbit, record the time $\tsp$, the splashback radius $\rrsp$, and the enclosed mass $\mmsp$. We exclude particles that were part of a subhalo larger than $0.01$ times the host mass at infall, because dynamical friction biases the $\rrsp$ of such particles. From the remaining distribution, we compute various estimators of $\rsp$ and $\msp$, namely, the mean, median, and higher percentiles of the distribution \citepalias{diemer_17_sparta}.

For the results presented in this paper, \sparta analyzed between $38$ and $640$ million particle apocenter passages per simulation, a total of $4.4$ billion splashbacks. Figure~\ref{fig:viz} shows a visualization of the conventional virial and splashback radii of halos with $\ntom \geq 1000$ particles (corresponding to $\mtom \geq 1.4 \times 10^{11} \msunh$) in a $30 \mpch$ slice through the L0125 simulation. The density field is visualized using the \gotetra code (Mansfield et al. 2017, in preparation), which is based on a tetrahedron density estimator \citep{abel_12, hahn_13, hahn_15}. The splashback radii shown correspond to $\rspes$, i.e., the radius enclosing $87\%$ of the particle apocenters (Table~\ref{table:defs}), the definition that most closely matches the results of \shellfish \citepalias{diemer_17_sparta}. Generally speaking, $\rsp$ is significantly larger than $\rvir$. A few halos were not assigned a splashback radius because they had recently been subhalos, but this fraction is relatively small (about $5\%$; see \citetalias{diemer_17_sparta}).


\section{Results}
\label{sec:results}

In this section, we analyze the distribution of $\rsp$, $\msp$, and $\deltasp$ as a function of halo mass, accretion rate, redshift, and cosmology. As shown in previous work \citep{diemer_14_profiles, more_15, mansfield_17}, the parameter that has the strongest influence on $\rsptom$ is the mass accretion rate. Thus, the majority of the section focuses on the \gammarsp. However, we also discuss the distribution of $\rsp$ marginalized over $\Gamma$, partly because the accretion rate of individual halos is difficult to measure observationally.

\subsection{Halo Sample}
\label{sec:results:sample}

As shown in \citetalias{diemer_17_sparta}, it does not matter which simulation a halo originated from because our results are insensitive to mass resolution as long as $\ntom \geq 1000$, a limit that is applied to all halo samples hereafter. We combine all halos with valid $\rsp$ and $\msp$ measurements into samples that are distinguished only by their redshift and cosmology. In order to compute $\Gamma$, we require halo masses at the current snapshot and at a particular time in the past. We exclude any halos that were not host halos at the current or past snapshot but include halos that temporarily became subhalos at intermediate times (so-called backsplash halos). We confirmed that excluding these halos makes a negligible difference to our results. We note that virtually all halos without a valid $\rsp$ measurement had recently been subhalos and might thus be excluded anyway.

Finally, we exclude the most extreme mass accretion rates from consideration. As discussed in \citetalias{diemer_17_sparta}, the lowest values of $\Gamma$ (in particular negative values) correspond to halos that are being disrupted because they are falling into or passing close by another halo. Due to the resulting tidal disruption, their radius and mass undergo drastic changes and are not particularly well defined, regardless of whether conventional definitions or $\rsp$ are used. Similarly, some halos are assigned very large values of $\Gamma$ that are indicative of a merger or disruption event. Thus, we exclude halos with $\Gamma < 0$ or $\Gamma > 12$ from our samples and do not include them when deriving our fitting function. This cut affects less than 1\% of halos at $z = 0$, and about 2\% at higher redshifts.

After all cuts, the sample for the fiducial cosmology includes about \num{250000} halos at $z = 0$, about \num{150000} at $z = 1$, and about $3500$ at $z = 8$. The $Planck$ sample contains about \num{170000} halos at $z = 0$ and about \num{120000} at $z = 1$. Unless stated otherwise, we plot the median $\rsp$ and $\msp$ of a halo sample because the mean is more sensitive to outliers. We compute the statistical uncertainty in each bin from the standard deviation and omit bins with fewer than 30 halos.

\subsection{Distribution and Scatter}
\label{sec:results:scatter}

\begin{figure}
\centering
\includegraphics[trim = 7mm 6mm 3mm 0mm, clip, scale=0.64]{\figdir/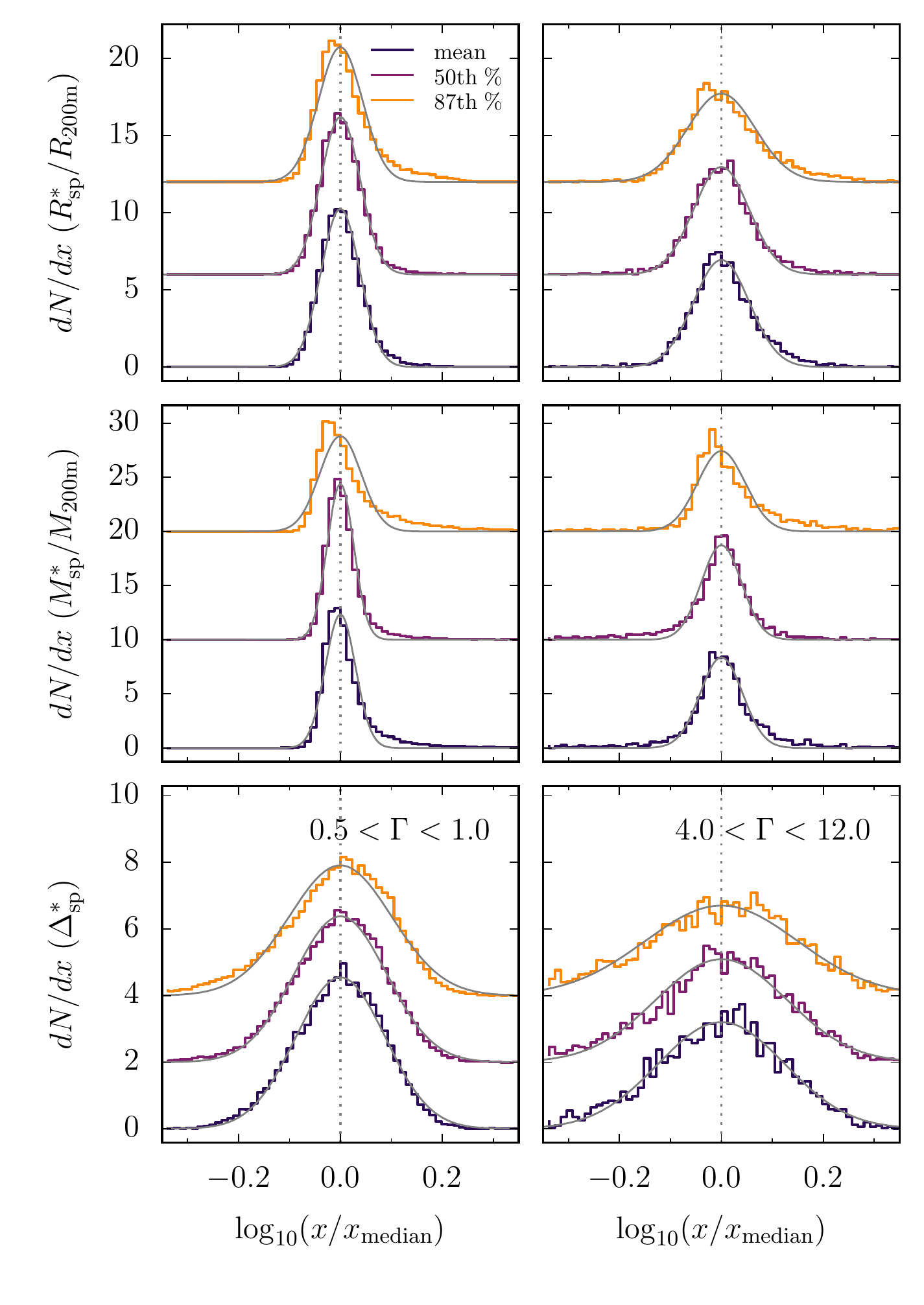}
\caption{Distribution of $\rsp$ (top row), $\msp$ (middle row), and $\deltasp$ (bottom row) for halos with $1 < \nu < 1.5$ at $z = 0.5$ (other samples exhibit similar distributions). The two columns refer to halos with moderate mass accretion rates (left) and very high accretion rates (right). Each colored line corresponds to a particular definition of $\rsp$, and the gray lines show the best-fit log-normal relations to those distributions (with a fixed median of $0$). The lines are offset from each other for clarity. The distributions are close to log-normal, though they exhibit tails toward high $\rsp$ and $\msp$ that are more prominent for higher percentiles. The width of the distribution increases with mass accretion rate and decreases with mass (see the text for a detailed discussion).}
\label{fig:distribution}
\end{figure}

We begin by analyzing the distributions of $\rsp$ and $\msp$ at a fixed mass accretion rate, mass, redshift, and cosmology. Figure~\ref{fig:distribution} shows examples of the distribution of residuals for three definitions of the splashback radius, mass, and overdensity, namely, $\rspmean$, $\rspmed$, and $\rspes$ (corresponding to the mean, median, and 87th percentile of the particle apocenter distribution). Generally, the distributions of $\rsp$ and $\msp$ are reasonably well described by log-normal functions, except for a tail toward positive values. The tails are stronger in $\msptom$ than in $\rsptom$, presumably because the splashback mass can increase due to large subhalos that have crossed into the halo but not yet influenced $\rsp$. The tails are weakest for $\rspmed$ and $\mspmed$, and increase toward higher percentiles. For example, the $\msptom$ distribution of the 87th percentile (orange lines in Figure~\ref{fig:distribution}) has a peak that is slightly shifted off the median value. The distribution of the enclosed overdensity $\deltasp$ is much wider due to the combined scatter from $\rsp$ and $\msp$, but shows no systematically discernible tails.

As the residuals from the median values are nearly log-normal, we will hereafter quantify the distributions as the median $\rsp$ or $\msp$ and the logarithmic 68\% scatter in dex. Figure~\ref{fig:distribution} hints at some of the most important trends: the scatter is smallest for low percentiles, low $\Gamma$, and large halo masses. In contrast, redshift does not have a major impact on the scatter (not shown in Figure~\ref{fig:distribution}). We find that the scatter, expressed in units of dex, can be approximated as 
\begin{equation}
\label{eq:sigma}
\sigma_{\rm sp} = \sigma_0 + \sigma_{\Gamma} \Gamma + \sigma_{\nu} \nu + \sigma_{\rm p} p
\end{equation}
where $p$ is the percentile divided by 100, and $\sigma_{\rm p}$ is zero for $\rspmean$ and $\mspmean$. The parameters differ slightly for $\rsp$ and $\msp$, and are given in Table~\ref{table:fits}. They were derived from a least-squares fit to the measured scatter in the \gammarsp of the fiducial and $Planck$ samples at redshifts $0.2$, $0.5$, $1$, $2$, $4$, and $8$ and in the peak-height bins shown in Figure~\ref{fig:fits} (we ignore the scatter at $z = 0$ which is artificially increased; see \citetalias{diemer_17_sparta}). The scatter in the enclosed overdensity $\deltasp$ is well approximated by the scatter in $\rsp$ and $\msp$ added in quadrature,
\begin{equation}
\label{eq:sigmadelta}
\sigma_{\Delta_{\rm sp}} = \sqrt{\sigma_{\msp}^2 + 3 \sigma_{\rsp}^2} \,.
\end{equation}
For example, the scatter at intermediate masses ($\nu = 1$) and accretion rates ($\Gamma = 1$) is about $0.045$ dex in both $\rspmean$ and $\mspmean$, and increases to about $0.055$ dex for the 87th percentile. The lowest scatter of about $0.02$ dex occurs at $\Gamma \approx 0.5$ and $\nu \approx 3$. We note that Equation~(\ref{eq:sigma}) extrapolates to lower (and even negative) scatter but should not be taken seriously below $\sigma = 0.02$. The highest scatter occurs at low masses ($\nu = 0.5$) and high accretion rates ($\Gamma = 10$), about $0.08$ dex for $\rspmean$ and $0.1$ dex for $\rspes$, resulting in a scatter of about $0.2$ dex in $\deltaspes$.

We note that Equation~(\ref{eq:sigma}) does not describe the scatter at $z = 0$, or, more generally, at the final redshift of a simulation. At those snapshots, the scatter is increased significantly by the correction term introduced to balance the asymmetric time distribution of particle splashbacks \citepalias{diemer_17_sparta}. This term de-biases the results, on average, but induces additional scatter that strongly depends on $\Gamma$ because the extrapolation in time is less reliable for rapidly evolving halos. In particular, the scatter is barely increased at low accretion rates ($\Gamma \lsim 1$) but increased by up to a factor of $2$ at high accretion rates. Finally, we caution that (due to the tails in the distributions) the 2$\sigma$ (i.e., 95\%) scatter can be slightly larger than twice the 1$\sigma$ (i.e. 68\%) scatter. The difference exhibits a rather complex dependence on mass and redshift, and we refrain from adding further complexity to our fitting function.

\subsection{Fitting Function}
\label{sec:results:fit}

\begin{deluxetable}{lcccc}
\tablecaption{Best-Fit parameters
\label{table:fits}}
\tablewidth{0pt}
\tablehead{
\colhead{Parameter} &
\colhead{$\rspmean$} &
\colhead{$\rsp^{\%}$} &
\colhead{$\mspmean$} &
\colhead{$\msp^{\%}$}
}
\startdata
\multicolumn{5}{c}{\rule{0pt}{2ex} Parameters for $\rsp$ and $\msp$} \\
\hline
\rule{0pt}{3ex} $a_0$                & $   0.6498$ & $   0.3203$ & $   0.6792$ & $   0.2648$ \\
\rule{0pt}{0pt} $b_0$                & $   0.6004$ & $   0.2674$ & $   0.4051$ & $   0.6660$ \\
\rule{0pt}{0pt} $b_{\Omega}$         & $   0.0920$ & $   0.1134$ & $   0.2919$ & $   0.1688$ \\
\rule{0pt}{0pt} $b_{\nu}$            & $   0.0616$ & $   0.2080$ & $0$         & $0$         \\
\rule{0pt}{0pt} $c_0$                & $  -0.8063$ & $  -0.9596$ & $   3.3659$ & $   4.7287$ \\
\rule{0pt}{0pt} $c_{\Omega}$         & $  17.5205$ & $  16.2459$ & $   1.4698$ & $   2.3889$ \\
\rule{0pt}{0pt} $c_{\nu}$            & $  -0.2935$ & $0$         & $  -0.0756$ & $  -0.0841$ \\
\rule{0pt}{0pt} $c_{\Omega 2}$       & $  -9.6243$ & $  -9.4979$ & $0$         & $0$         \\
\rule{0pt}{0pt} $c_{\nu 2}$          & $   0.0392$ & $  -0.0185$ & $0$         & $0$         \\
\multicolumn{5}{c}{\rule{0pt}{2ex} Meta-Parameters for Dependence on Percentile} \\
\hline
\rule{0pt}{3ex} $a_{\rm p}$          & $0$         & $   0.6148$ & $0$         & $   0.8435$ \\
\rule{0pt}{0pt} $b_{\rm p}$          & $0$         & $   0.5452$ & $0$         & $  -0.6392$ \\
\rule{0pt}{0pt} $b_{\Omega \rm p}$   & $0$         & $0$         & $0$         & $   0.0032$ \\
\rule{0pt}{0pt} $b_{\Omega \rm p2}$  & $0$         & $0$         & $0$         & $   4.9393$ \\
\rule{0pt}{0pt} $b_{\nu \rm p}$      & $0$         & $  -0.2233$ & $0$         & $   0.2254$ \\
\rule{0pt}{0pt} $c_{\Omega \rm p}$   & $0$         & $   0.0039$ & $0$         & $  -0.7057$ \\
\rule{0pt}{0pt} $c_{\Omega \rm p2}$  & $0$         & $   8.9691$ & $0$         & $  -1.2419$ \\
\rule{0pt}{0pt} $c_{\Omega \rm 2p}$  & $0$         & $  -0.0005$ & $0$         & $0$         \\
\rule{0pt}{0pt} $c_{\Omega \rm 2p2}$ & $0$         & $  10.6132$ & $0$         & $0$         \\
\rule{0pt}{0pt} $c_{\nu \rm p}$      & $0$         & $  -0.4511$ & $0$         & $  -0.3911$ \\
\rule{0pt}{0pt} $c_{\nu \rm 2p}$     & $0$         & $   0.0880$ & $0$         & $   0.0742$ \\
\multicolumn{5}{c}{\rule{0pt}{2ex} Parameters for 68\% Scatter (in dex)} \\
\hline
\rule{0pt}{3ex} $\sigma_0$           & $   0.0526$ & $   0.0445$ & $   0.0528$ & $   0.0276$ \\
\rule{0pt}{0pt} $\sigma_\Gamma$      & $   0.0038$ & $   0.0044$ & $   0.0025$ & $   0.0023$ \\
\rule{0pt}{0pt} $\sigma_\nu$         & $  -0.0121$ & $  -0.0146$ & $  -0.0112$ & $  -0.0125$ \\
\rule{0pt}{0pt} $\sigma_{\rm p}$     & $0$         & $   0.0226$ & $0$         & $   0.0473$
\enddata
\end{deluxetable}

\begin{figure*}
\centering
\includegraphics[trim = 2mm 6mm 3mm 0mm, clip, scale=0.56]{\figdir/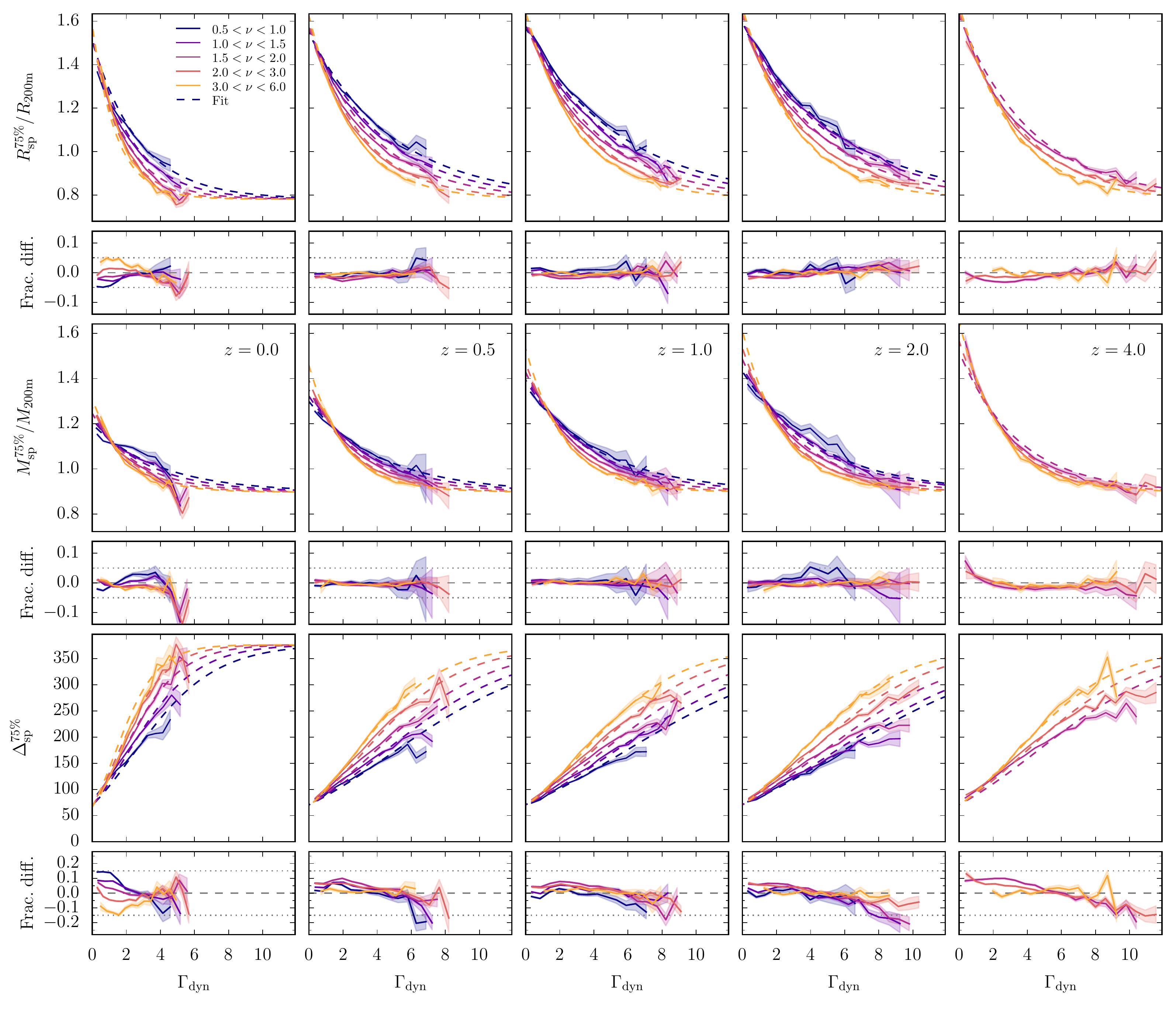}
\caption{Relations between $\Gamma$ and $\rsp$ (top two rows), $\msp$ (middle two rows), and $\deltasp$ (bottom two rows), at different redshifts (columns) and peak heights (colors). The solid lines show the median relations, and the shaded areas show the statistical uncertainty. The dashed lines correspond to the fitting function described in Section~\ref{sec:results:fit}. The smaller panels in each set of rows show the differentials between the fits and simulation data. The relations shown are for the fiducial cosmology and the 75th percentile, but the results for other definitions and the $Planck$ cosmology are equally well fit.}
\label{fig:fits}
\end{figure*}

Before we discuss the various dependencies of $\rsp$ and $\msp$ in detail, we summarize our results with a convenient fitting function. We find that the $\Gamma$--$\rsptom$ and $\Gamma$--$\msptom$ relations are --- at any redshift, cosmology, peak height, and for any $\rsp$ definition --- well fit by an expression similar to those suggested by \citet{diemer_14_profiles} and \citet{more_15},
\begin{equation}
\label{eq:fit1}
\xsp = A + B e^{-\Gamma/C}
\end{equation}
where $\xsp$ can stand for either $\rsp$ or $\msp$, and $A$, $B$, and $C$ are free parameters. Those parameters are, in turn, functions of mass and redshift such that
\begin{align}
\label{eq:fit2}
A & = A_0 \nonumber \\
B & = (B_0 + B_{\Omega} \Omega_{\rm m}) \times (1 + B_{\nu} \nu) \nonumber \\
C & = (C_0 + C_{\Omega} \Omega_{\rm m} + C_{\Omega 2} \Omega_{\rm m}^2) \times (1 + C_{\nu} \nu + C_{\nu 2} \nu^2)
\end{align}
where we have introduced a total of nine free parameters. However, not all of these parameters are necessary to fit either $\rsp$ or $\msp$. In principle, there is no reason to expect that $\rsp$ and $\msp$ should be fit by exactly the same functional form. Thus, it is not surprising that slightly different parameters are used in the two fits. Furthermore, the fit parameters in Equation~(\ref{eq:fit2}) depend on the definition of $\rsp$. We fit $\rspmean$ and $\mspmean$ separately and opt to further parameterize the dependence of the fit parameters on the percentile. For this purpose, we introduce $p$, the percentile value divided by $100$ (e.g., $0.5$ for the median). The fit parameters depend on $p$ in a nontrivial manner, where
\begin{align}
\label{eq:fit3}
A_0 &= a_{0} + a_{\rm p} \times p \nonumber \\
B_0 &= b_{0} + b_{\rm p} \times p \nonumber \\
B_{\Omega} &= b_{\Omega} + b_{\Omega \rm p} \times \exp( b_{\Omega \rm p2} \times p) \nonumber \\
B_{\nu} &= b_{\nu} + b_{\nu \rm p} \times p	 \nonumber \\
C_0 & = c_0 \nonumber \\
C_{\Omega} &= c_{\Omega} + c_{\Omega \rm p} \times \exp(c_{\Omega \rm p2} \times p)		 \nonumber \\
C_{\Omega 2} &= c_{\Omega 2} + c_{\Omega \rm 2p} \times \exp(c_{\Omega \rm 2p2} \times p) \nonumber \\	
C_{\nu} &= c_{\nu} + c_{\nu \rm p} \times p \nonumber \\
C_{\nu 2} &= c_{\nu 2} + c_{\nu \rm 2p} \times p	\,.
\end{align}
We constrain all free parameters simultaneously using a Levenberg--Marquart least-squares fit to the median \gammarsp at redshifts $0$, $0.2$, $0.5$, $1$, $2$, $4$, and $8$, at the same peak-height bins as those shown in Figure~\ref{fig:fits}, and in both the fiducial and $Planck$ cosmologies. The statistical uncertainty is used as an inverse weight in the fit. However, in this scheme, low $\Gamma$ values are weighted much more heavily than high $\Gamma$ values where the halo sample is less populated, and the $\chi^2$ values are much greater than one, indicating that the error bars are underestimated. Thus, we add a $1\%$ systematic error in quadrature with the statistical error, which balances the weights and leads to more reasonable $\chi^2$ values between $1$ and $2$. We constrain the dependence of the parameters on $p$ by simultaneously fitting the 50th, 63rd, 75th, and 87th percentiles. Based on the results of \citetalias{diemer_17_sparta}, the highest percentiles are known to be unreliable, and we thus do not attempt to fit their \gammarsp. The fitting function should not be extrapolated beyond the 50th and 87th percentiles or beyond the range of mass accretion rates for which it was constrained ($0 < \Gamma < 12$).

Figure~\ref{fig:fits} shows a summary of our main results and fitting function. Each column shows $\rsp$, $\msp$, and $\deltasp$ for a different redshift, and the colors indicate different halo masses. Given the statistical uncertainties, the fits agree with the median relations to 5\% or better everywhere in $\Gamma$--$\Omega_{\rm m}$--$\nu$ parameter space where data are available. Figure~\ref{fig:fits} shows $\rspsf$, but the same holds for $\rspmean$ and up to the 87th percentile. Similarly, Figure~\ref{fig:fits} shows the results for our fiducial cosmology, but the $Planck$ results are fit equally well. We note that the relations for the median $\rsp$ and $\msp$ do not necessarily have to predict the correct median enclosed overdensity $\deltasp$ if the distributions are asymmetric. However, combining the fitting functions for $\rsp$ and $\msp$, we find that the result has roughly the expected agreement (bottom row of Figure~\ref{fig:fits}; any error on $\rsp$ is tripled; see Equation~(\ref{eq:sigmadelta})). We have implemented our fitting function in the publicly available Python code \colossus\footnote{\colossus is a Python module for computations related to cosmology, large-scale structure, and dark matter halos \citep{code_colossus, diemer_15_cm}. In addition to all fitting functions proposed in this paper, we have also implemented the \citet{more_15} fit as well as the \citet{adhikari_14} and \citet{shi_16} semi-analytical models (see Section~\ref{sec:comp}).}.

Our fitting function highlights a number of interesting features in the data. First, \citet{more_15} found no significant evidence for a dependence of $\rsp$ on halo mass, largely due to the limited accuracy of the $\rsp$ determination from stacked density profiles. Here, we qualitatively confirm the results of \citet{mansfield_17} in that we find a slight but significant dependence on mass (or, equivalently, $\nu$). Interestingly, this mass dependence is weak for $\rspmean$ and $\rspmed$, but becomes more significant at higher percentiles. We discuss the mass dependence further in Section~\ref{sec:discussion:universality}. 

According to our fitting function, $\rsptom$ and $\msptom$ approach constant values at high $\Gamma$ that do not depend on redshift, mass, or cosmology (they do, however, depend on the percentile definition). We caution that our data do not unambiguously require such behavior. But the data also do not show a statistically significant preference for an evolution of the high-$\Gamma$ value with mass or redshift. The fits for $\rsp$ and $\msp$ imply that $\deltasp$ also asymptotes to a particular value at high $\Gamma$, about $500$ for $\rspmean$. At $z = 0$ and for our fiducial cosmology, $\Delta_{\rm m} = 500$ corresponds to $\Delta_{\rm c} = 135$, meaning that the average $\rsp$ (using any definition) is always larger than $\rtoc$, even at very high accretion rates. At higher redshift, however, $\Delta_{\rm c}$ becomes similar to $\Delta_{\rm m}$, meaning that $\rsp$ can reach radii smaller than $\rtoc$. 

\vspace{1cm}

\subsection{Dependence on Redshift and Cosmology}
\label{sec:results:redshift}

\begin{figure}
\centering
\includegraphics[trim = 3mm 6mm 3mm 0mm, clip, scale=0.62]{\figdir/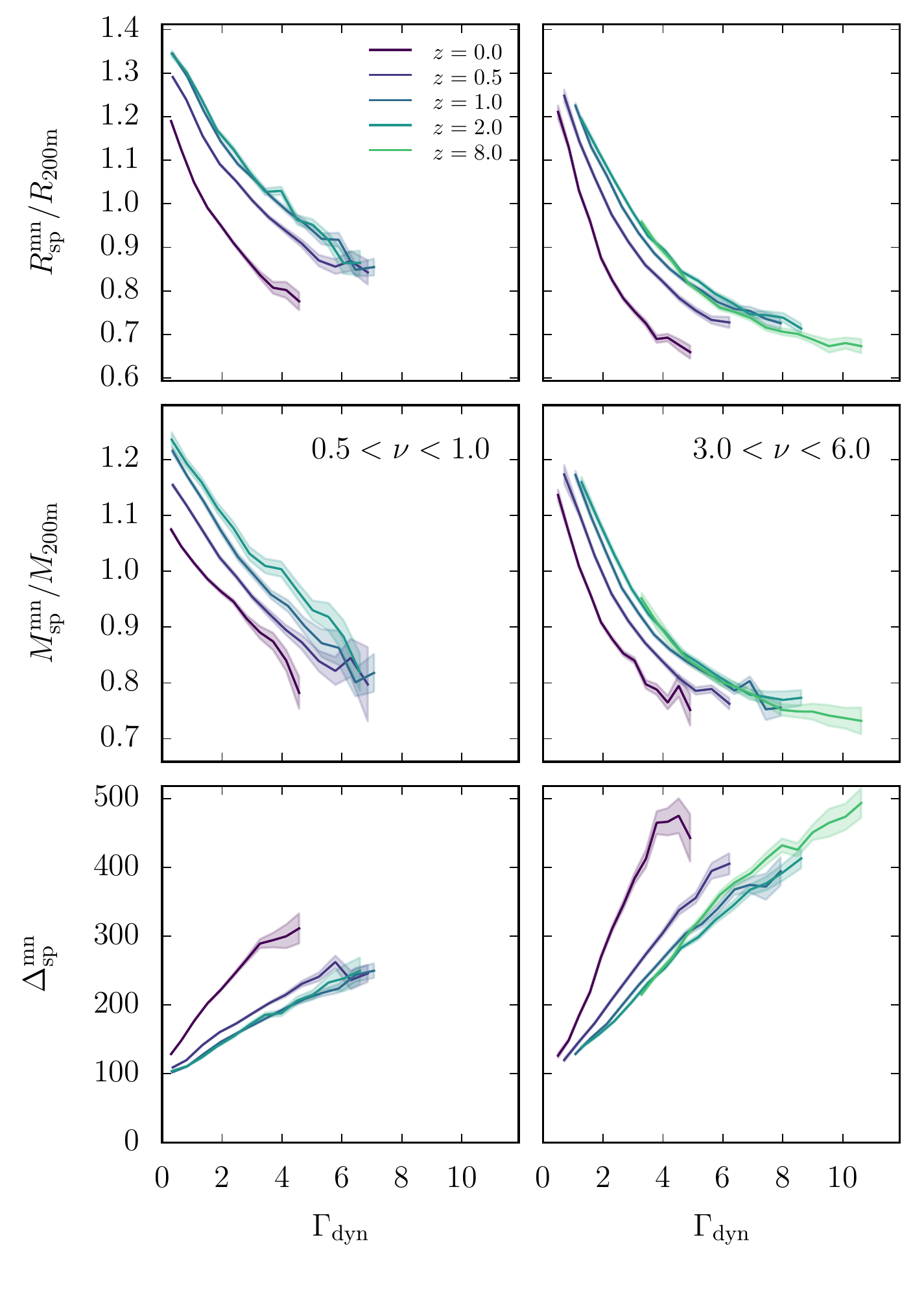}
\caption{Dependence of the \gammarsp on redshift for low-mass (left) and high-mass (right) halos. The evolution supports the notion that $\rsp$ depends on $\Omega_{\rm m}$ rather than redshift, as the relations converge at $z \gsim 2$ where $\Omega_{\rm m} \approx 1$. }
\label{fig:redshift}
\end{figure}

\begin{figure}
\centering
\includegraphics[trim = 3mm 6mm 3mm 0mm, clip, scale=0.62]{\figdir/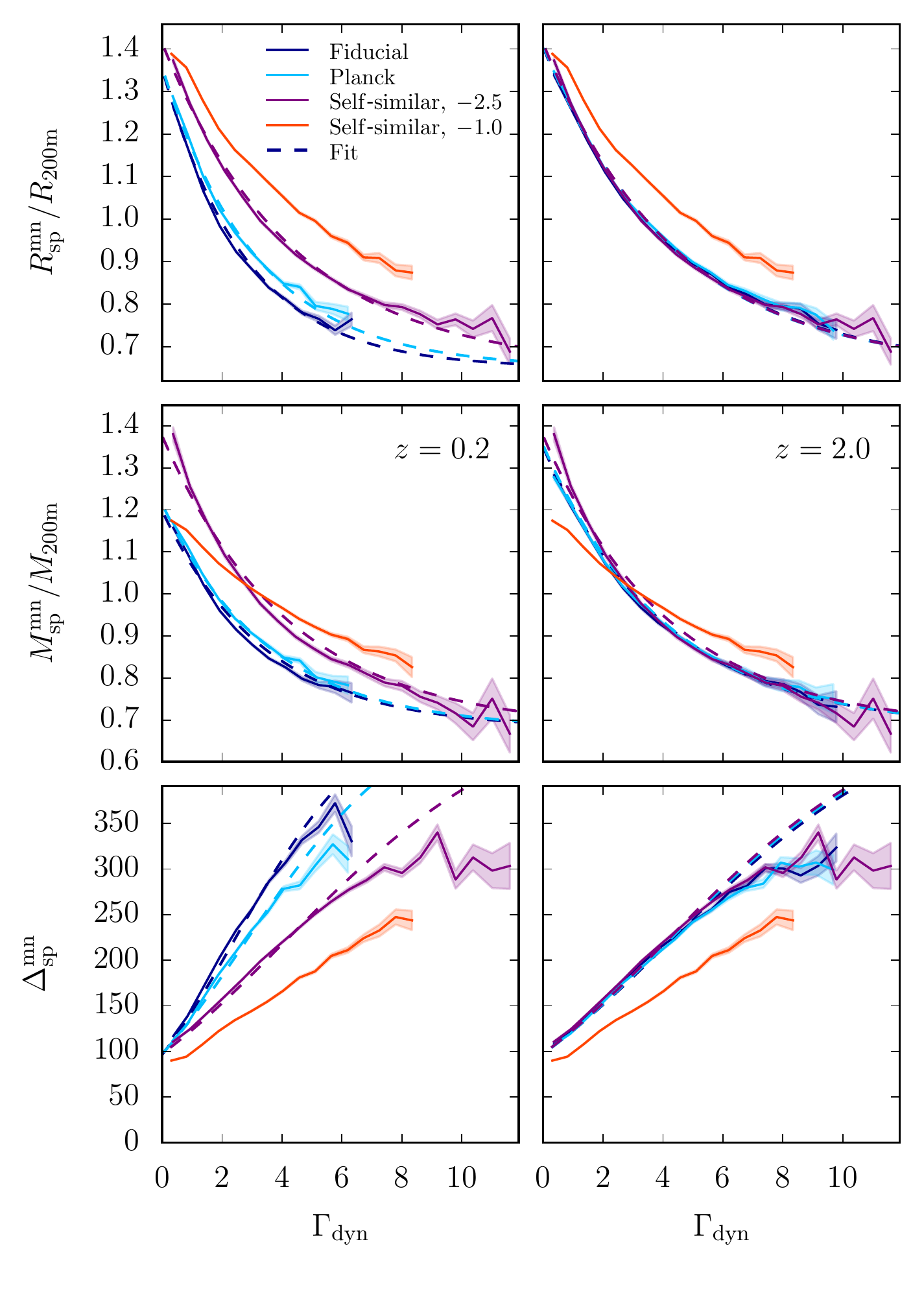}
\caption{Dependence of the \gammarsp on cosmology. The dark blue lines correspond to the fiducial cosmology, the light blue lines to the $Planck$ cosmology, and the purple and red lines to the self-similar cosmologies with power spectrum slopes of $n = -2.5$ and $-1$, respectively. At low redshift (left column), $\Omega_{\rm m}$ differs between the fiducial and $Planck$ cosmologies ($0.27$ and $0.32$, respectively), leading to slightly different $\rsptom$ and $\msptom$. At higher redshifts, where $\Omega_{\rm m} \approx 1$ in both cosmologies, the relations are indistinguishable (right column). These effects are correctly captured by our fitting function (dashed lines). The self-similar cosmologies have $\Omega_{\rm m} = 1$ at all times, meaning the respective relations are the same in the left and right columns. While they are clearly distinct from the fiducial and $Planck$ cosmologies at low redshift, the $n = -2.5$ cosmology is very similar at high redshift. A power-spectrum slope of $n = -1$ is much shallower than the $\Lambda$CDM power spectrum, leading to significantly different relations at all redshifts. This difference demonstrates that cosmological parameters other than $\Omega_{\rm m}$ can have an impact on $\rsp$.}
\label{fig:cosmo}
\end{figure}

\begin{figure}
\centering
\includegraphics[trim = 3mm 6mm 3mm 0mm, clip, scale=0.62]{\figdir/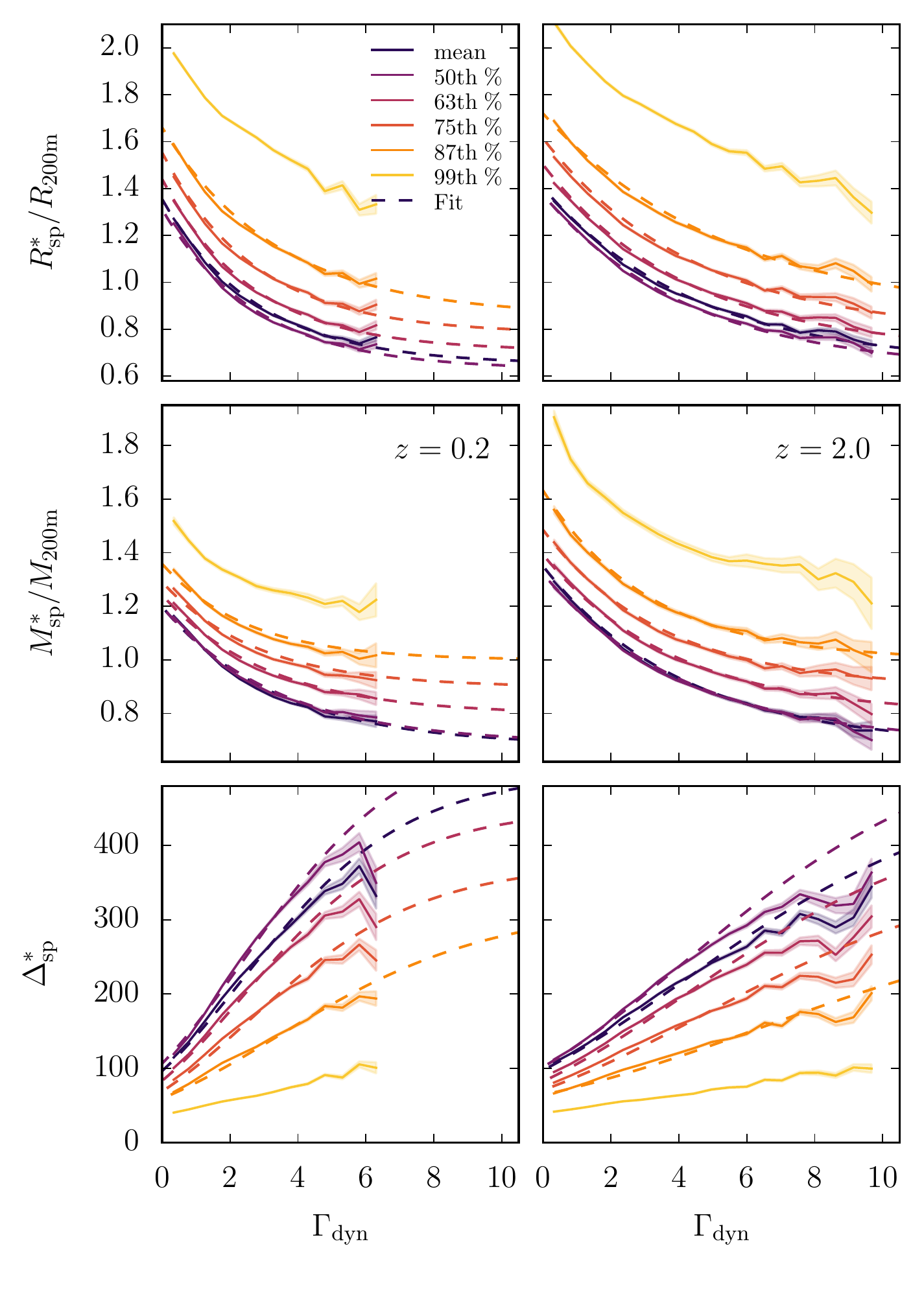}
\caption{The \gammarsp for different definitions of $\rsp$, for halos with $1.5 < \nu < 2$ at redshifts $0.2$ (left column) and $2.0$ (right column). Our fitting function (dashed lines) captures the differences up to the 87th percentile. As expected, the mean and median (dark blue and purple lines) are similar, but higher percentiles lead to increasingly higher values of $\rsp$.}
\label{fig:stats}
\end{figure}

Another noteworthy feature of the fitting function presented in Equations~(\ref{eq:fit1}--\ref{eq:fit3}) is that it encapsulates any dependence on redshift and cosmology into a dependence on $\Omega_{\rm m}(z)$. Figure~\ref{fig:redshift} demonstrates that such a scaling is strongly suggested by the data. Instead of comparing different peak heights within each panel as in Figure~\ref{fig:fits}, the columns show different peak heights, and the lines in each panel correspond to different redshifts. At redshifts higher than $z \approx 2$, $\Omega_{\rm m}$ barely changes, which manifests itself as a constant \gammarsp, even at $z = 8$. We find similar dependencies for other peak-height bins, as well as for higher percentiles. 

Although similar redshift scalings were predicted in analytical models \citep{adhikari_14, shi_16} and seemed to work well for the fitting functions of \citet{more_15} and \citet{mansfield_17}, the dependence on cosmology has not been explicitly tested in simulations. Figure~\ref{fig:cosmo} compares the \gammarsp for a particular peak-height bin in a number of cosmologies. First, we focus on the fiducial and $Planck$ cosmologies (dark and light blue lines, respectively). As expected, the $Planck$ cosmology produces slightly higher values of $\rsptom$ at low redshifts due to its higher $\Omega_{\rm m,0} = 0.32$ (compared to $0.27$ in the fiducial cosmology). At high redshift, the difference vanishes, indicating that a scaling with $\Omega_{\rm m}$ captures the difference between the cosmologies. We note that the $Planck$ cosmology also has a $\sigma_8$ that is 2\% higher than that in the fiducial cosmology, but our data are not constraining enough to definitively exclude dependencies on $\sigma_8$ or other cosmological parameters.

We further test the cosmology dependence using simulations that have $\Omega_{\rm m} = 1$ throughout, namely, self-similar cosmologies with a power-law power spectrum (Table~\ref{table:cosmo}). In such universes, we expect the \gammarsp not to evolve with redshift at all. We confirm this self-similar scaling, which constitutes further evidence that $\rsp$ depends on $\Omega_{\rm m}$ rather than $z$. Furthermore, the self-similarity allows us to combine halos at different redshifts into one sample per simulation, ensuring coverage over a wide range of peak heights. The samples from each self-similar simulation differ only by the slope of the initial power spectrum, $n$ (see \citealt{diemer_15_cm} for more details on the simulations and the technique of combining redshifts). Figure~\ref{fig:cosmo} shows the \gammarsp in two self-similar cosmologies, namely, those with $n = -2.5$ (purple) and $n = -1$ (red). A slope of $-2.5$ is close to the slope on scales relevant for halo formation in a $\Lambda$CDM cosmology \citep[e.g., Figure 4 in][]{diemer_15_cm}. Our fitting function (with $\Omega_{\rm m} = 1$) matches the \gammarsp from this simulation well, even though the self-similar models were not used when constraining the fit parameters. Nevertheless, the self-similar cosmology with $n = -2.5$ exhibits essentially the same relations as our fiducial cosmology at high redshift. This match is yet another confirmation that $\Omega_{\rm m}$ is the variable that controls $\rsp$, not redshift. 

By contrast, the self-similar simulation with a much shallower power spectrum slope, $n = -1$, leads to a rather different \gammarsp that is not described by our fitting function. We conclude that $\Omega_{\rm m}$ is not the only parameter than influences $\rsp$; the power-spectrum slope clearly has an impact too. In principle, we could introduce $n$ as an extra parameter into our fitting function and use the self-similar models to constrain the dependence of $\rsp$. However, the impact of $n$ in $\Lambda$CDM is degenerate with the effects of $\Omega_{\rm m}$ and mass, making it difficult to disentangle the dependencies. Furthermore, it is not clear a priori how to define $n$ in a $\Lambda$CDM cosmology where the slope is scale-dependent (and thus halo mass--dependent). As the self-similar models have little practical application, we leave an investigation of the effect of $n$ for future work.

\subsection{Definitions of $\rsp$}
\label{sec:results:stat}

Besides $\nu$ and $\Omega_{\rm m}$, our fitting function depends on the definition of $\rsp$ and $\msp$, i.e. whether we use the mean of the $\rrsp$ and $\mmsp$ distributions, their median, or higher percentiles. In Section~\ref{sec:discussion}, we will demonstrate that different definitions can be useful for different purposes. Figure~\ref{fig:stats} shows the \gammarsp for different definitions of $\rsp$. At first sight, it appears that the main difference is the normalization of the curves. However, there are also nontrivial changes in the shape of the relations, demanding the relatively large number of free parameters introduced in Equation~(\ref{eq:fit3}). At the highest percentiles (greater than the 87th), the evolution of the normalization becomes superlinear, and the shape changes in a complex manner with percentile. As the measurement of the highest percentiles is relatively uncertain anyway \citepalias{diemer_17_sparta}, we limit the applicability of our fitting function to the range between the 50th and 87th percentiles and omit it for the 99th percentile in Figure~\ref{fig:stats}. 

\subsection{Constraints on $\rsp$ without Knowledge of the Mass Accretion Rate}
\label{sec:results:nogamma}

\begin{figure}
\centering
\includegraphics[trim = 8mm 6mm 3mm 0mm, clip, scale=0.67]{\figdir/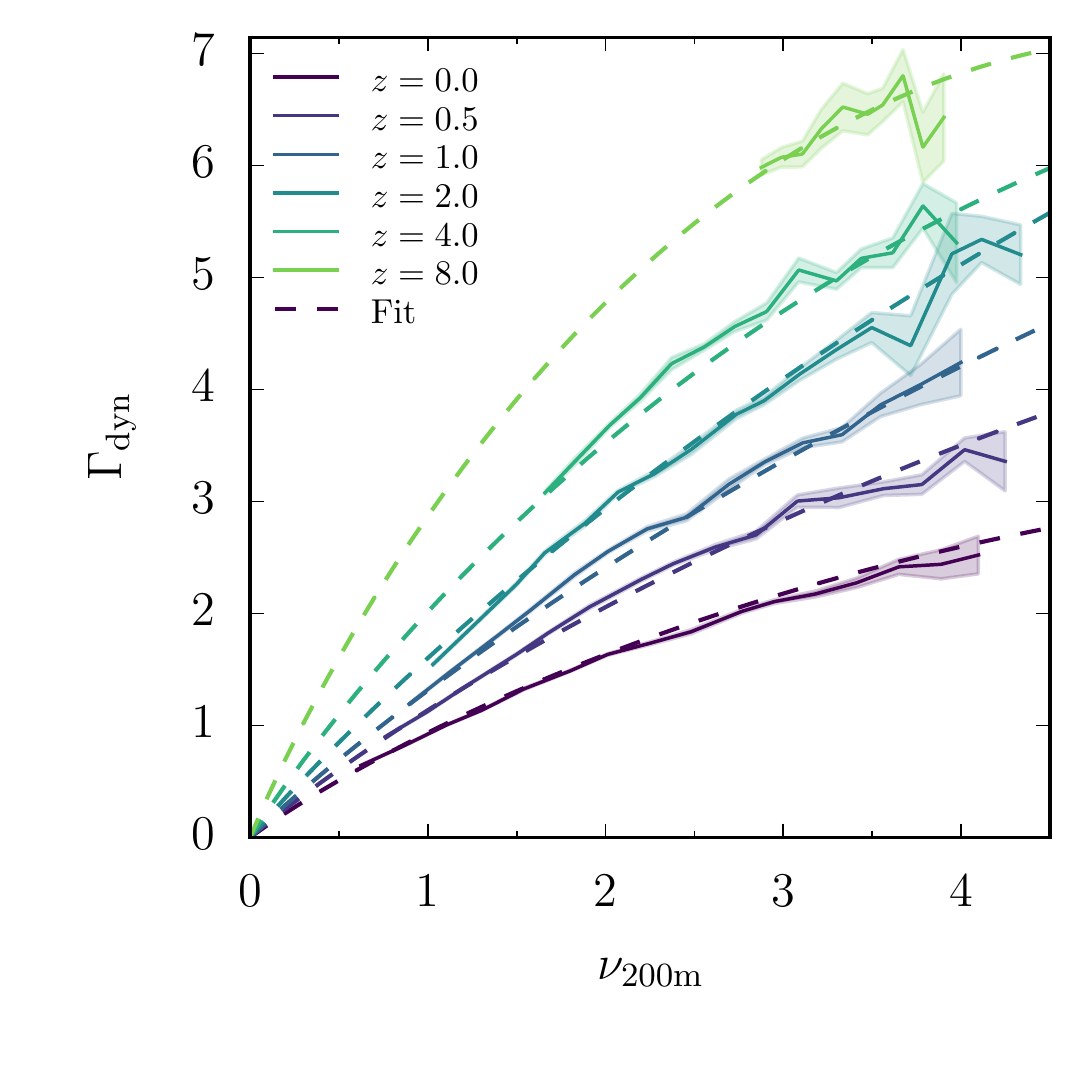}
\caption{Mass accretion rate as a function of peak height and redshift for the fiducial cosmology. The solid lines show the median $\Gamma$ at a given $\nu$ and $z$, the shaded areas show the statistical uncertainty, and the dashed lines show the fitting function given in Equation~(\ref{eq:gamma_fit}). We note that the 68\% scatter (not shown) in the relations is large: between about $0.15$ and $0.35$ dex, depending on redshift and halo mass.}
\label{fig:nugamma}
\end{figure}

\begin{figure}
\centering
\includegraphics[trim = 3mm 6mm 3mm 0mm, clip, scale=0.62]{\figdir/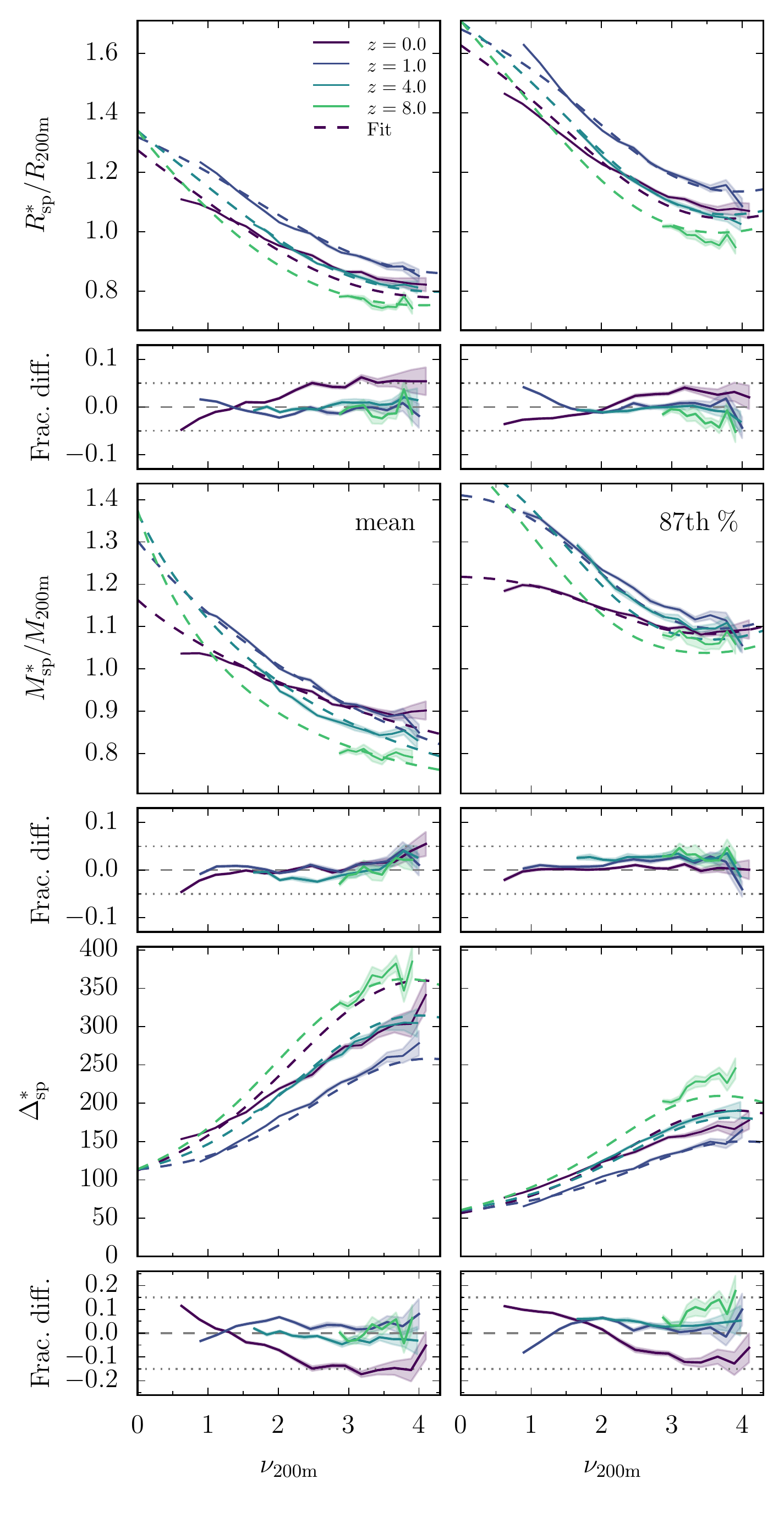}
\caption{Splashback radius (top), mass (middle), and overdensity (bottom) as a function of peak height, marginalized over all mass accretion rates. The solid lines and shaded areas show the median relations and statistical uncertainties for different redshifts, as well as for the mean (left column) and 87th percentile (right column) definitions. The dashed lines show the fit to the \gammarsp (Equations~(\ref{eq:fit1}--\ref{eq:fit3})), evaluated with $\Gamma$ calculated from the fit to the \nugamma (Equation~(\ref{eq:gamma_fit})). The evolution of the relations with redshift is nontrivial due to the competing effects of an increasing $\Omega_{\rm m}$ and increasing $\Gamma$ at high redshift (see text for a detailed discussion).}
\label{fig:nuqsp}
\end{figure}

So far, we have considered $\rsp$ primarily as a function of $\Gamma$ and secondarily of other variables because $\Gamma$ has the strongest effect. Unfortunately, $\Gamma$ is also the variable that is hardest to measure observationally: in practice, we almost never know the mass accretion rate of individual halos. Thus, we also give expressions for $\rsp$ in the absence of any knowledge about $\Gamma$, i.e. the median $\rsp$ as a function of halo mass and redshift but marginalized over all mass accretion rates. 

First, it is instructive to consider the average $\Gamma$ as a function of halo mass and redshift, as shown in Figure~\ref{fig:nugamma}. The accretion rate increases with peak height (because larger halos are more likely to be actively forming at any given time in hierarchical structure formation), but there is also a strong trend with redshift, with much higher $\Gamma$ at high redshift. Moreover, the distribution at fixed mass and redshift is broad and not particularly well described by a normal or log-normal distribution (partially because a few percent of halos have negative accretion rates, especially at low $\nu$). Neglecting a mild dependence on redshift, we find that the logarithmic scatter in the $\Gamma$ distributions is roughly
\begin{equation}
\label{eq:gamma_scatter}
\sigma_{\Gamma} \approx 0.41 - 0.07 \nu
\end{equation}
where $\sigma_{\Gamma}$ is in units of dex. Despite the large scatter, there are clear trends in the median accretion rate $\Gamma(\nu, z)$ that can be approximated with the expression
\begin{equation}
\gammadyn = A \nu + B \nu^{3/2}
\end{equation}
where 
\begin{align}
\label{eq:gamma_fit}
A &= 1.2222 + 0.3515 z \nonumber \\
B &= -0.2864 + 0.0778 z - 0.0562 z^2 + 0.0041 z^3 \,.
\end{align}
This fitting function is shown with dashed lines in Figure~\ref{fig:nugamma}, and fits the median $\Gamma$ to better than 5\% at all $\nu$ and $z$ and for both the fiducial and $Planck$ cosmologies. It is clear that a dependence on $\Omega_{\rm m}$ instead of $z$ would not work in this case, as $\Gamma$ strongly increases at high redshift even though $\Omega_{\rm m} \approx 1$ \citep[e.g.,][]{zhao_09_mah}.

Given the trends in $\Gamma(\nu, z)$, we expect the \nursp to experience a conflation of multiple competing effects: the \gammarsp increases with redshift due to an increasing $\Omega_{\rm m}$, but the increasing $\Gamma$ at high $z$ leads to lower $\rsp$. As the distribution of $\Gamma$ is nonsymmetric and the \gammarsp is nonlinear, there is no guarantee that we can construct a \nursp from our $\nu$-$\Gamma$ and \gammarsps. However, we find that such a procedure does, in fact, work surprisingly well. Figure~\ref{fig:nuqsp} shows the \nursp for a number of redshift bins and $\rsp$ definitions. The dashed lines show the fit obtained from our $\Gamma$-based fitting function (Equations~(\ref{eq:fit1}--\ref{eq:fit3})), evaluated using the $\Gamma$ fit of Equation~(\ref{eq:gamma_fit}). The fits are accurate to 5\% for $\rsptom$ and $\msptom$ at all peak heights, redshifts, and $\rsp$ definitions and for both the fiducial and $Planck$ cosmologies. As expected, the corresponding maximum error in $\deltasp$ is about 15\%. 

Naturally, the scatter in the \nursp is increased compared to that of the \gammarsp because we are averaging over all mass accretion rates. In particular, the 68\% scatter is about $0.07$ dex in $\rsptom$ regardless of peak height or redshift and between $0.04$ dex and $0.1$ dex in $\msptom$, where the highest scatter occurs at high redshift and low peak height. The distributions in $\rsptom$ and $\msptom$ combine to a more or less constant scatter of $0.15$ dex in $\deltasp$. As with the \gammarsp, the distribution of $\rsp$ and $\msp$ values is reasonably described by a log-normal, but the tails toward high and low values are enhanced when marginalizing over $\Gamma$. As a result, the 2$\sigma$ (i.e., 95\%) scatter can be slightly larger than twice the 1$\sigma$ (i.e., 68\%) scatter. The exact distribution shows complex dependencies on peak height and redshift, and we refrain from quantifying it further. 

Overall, the $\sim 0.07$ dex scatter in the \nursp is surprisingly low, considering the scatter in the \gammarsp and that the distribution of accretion rates is relatively extended. The low scatter allows for meaningful inferences regarding $\rsp$ and $\msp$ in the absence of any knowledge about $\Gamma$. 


\section{Comparison to Previous Work}
\label{sec:comp}

The \sparta algorithm operates based on an entirely different principle than any of the previous estimates of $\rsp$ used in either simulations or observations. Thus, we expect that our findings might disagree with other measurements and models. In this section, we compare our data with previous simulation results based on density profiles and the \shellfish algorithm, as well as with theoretically motivated models.

\subsection{Comparison with Results Based on Density Profiles}
\label{sec:comp:profiles}

\begin{figure}
\centering
\includegraphics[trim = 3mm 6mm 3mm 0mm, clip, scale=0.62]{\figdir/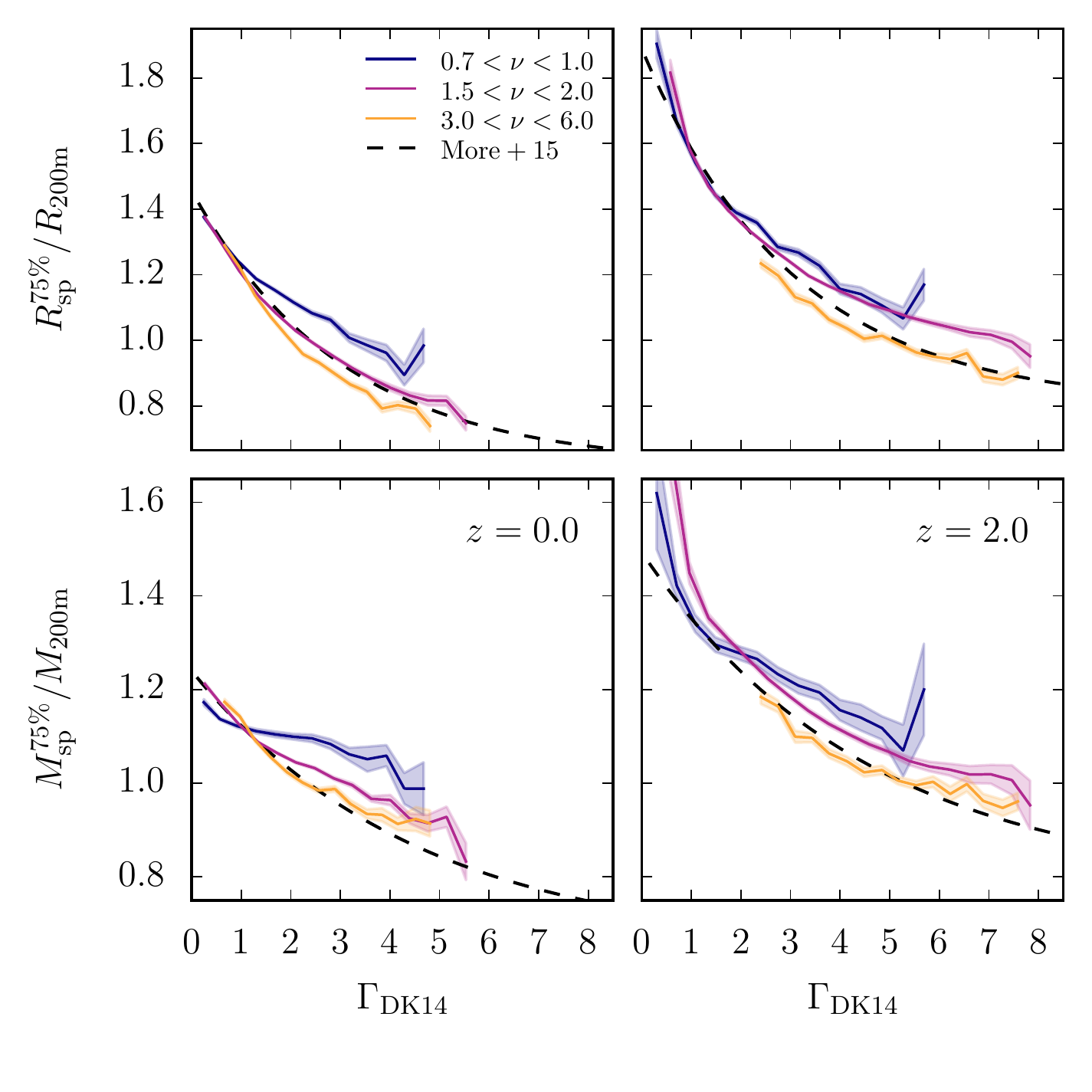}
\caption{Comparison of the median $\rsptom$ (top) and $\msptom$ (bottom) with the fitting function of \citet[][dashed lines]{more_15}, where $\rsp$ and $\msp$ are defined as the $75$th percentile of the distribution of individual particles' $\rrsp$ and $\mmsp$. The formula of \citet{more_15} was calibrated from stacked halo density profiles. For compatibility, we express the mass accretion rate as $\gammadk$ in this figure. The \citet{more_15} formula does not take the dependence on peak height into account but roughly matches the \sparta results for the highest peak-height bin. This match indicates that the radius of the steepest density slope includes about 75\% of the particle apocenters.}
\label{fig:more15_gamma}
\end{figure}

Most work on the splashback radius thus far has been based on spherically averaged density profiles, both in simulations and in observations. As those profiles suffer from noise due to resolution and substructure effects, they cannot generally be used to measure $\rsp$ for individual halos. Thus, \citet{more_15} used the stacked density profiles of \citet{diemer_14_profiles} and defined $\rsp$ as the radius where the logarithmic slope of the median profile is steepest. Based on this definition, they found that $\rsp$ decreases as a function of $\Gamma$ and increases slightly with increasing $\Omega_{\rm m}$ --- trends which we confirm in this work. Presumably owing to the relatively poor accuracy and restricted range of peak height used by \citet{more_15}, they did not detect any dependence on mass at fixed $\Gamma$, as found in this investigation and in \citet{mansfield_17}.

With the data at hand, we can for the first time elucidate the relation between the radius of the steepest slope and the apocenter passages of particles. Figure~\ref{fig:more15_gamma} compares the \gammarsp as derived by \sparta to the \citet{more_15} fitting function based on stacked density profiles. We note that the mass accretion rates $\gammadk$ were computed based on $\mvir$ in \citet{more_15} and based on $\mtom$ in this work, but the difference is negligible given the accuracy of this comparison. We choose $\rspsf$ as this definition matches the \citet{more_15} results most closely, indicating that the steepest profile slope occurs at a radius that encloses about 75\% of the particle apocenters. The \citet{more_15} function matches the overall shape and redshift evolution of the relations relatively well, particularly for high-mass halos. Choosing a lower percentile might bring the overall normalizations into better agreement at low mass, but the \sparta relations become almost mass-independent at low $\Gamma$. Thus, there is no percentile for which the low-mass relation is matched well by the fit of \citet{more_15}. We further discuss the connection between the density and splashback profiles in Section~\ref{sec:discussion:universality}. 

\begin{figure}
\centering
\includegraphics[trim = 3mm 6mm 3mm 0mm, clip, scale=0.62]{\figdir/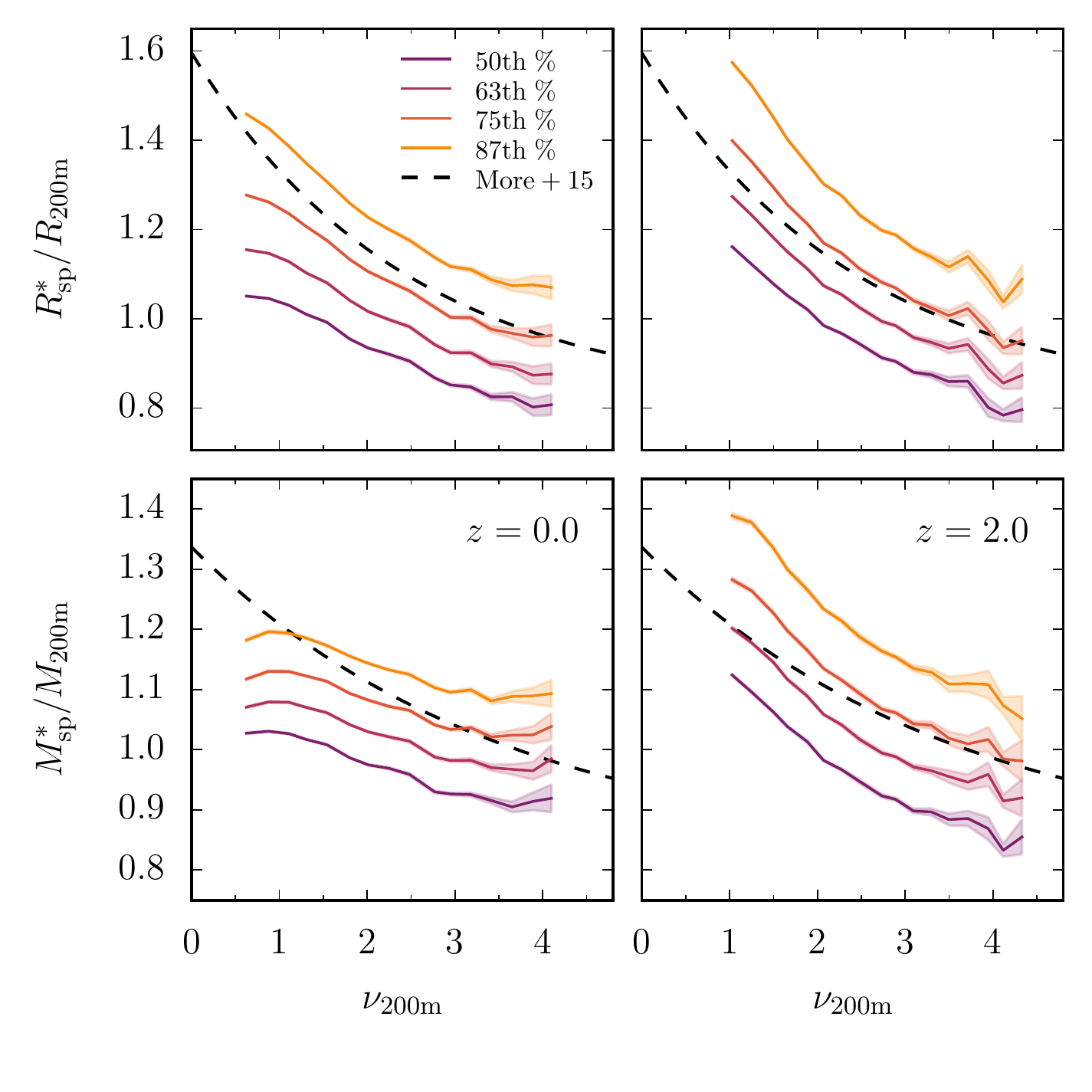}
\caption{Same as Figure~\ref{fig:more15_gamma}, but as a function of $\nu$ rather than $\Gamma$. The colored lines show different $\rsp$ definitions. As in Figure~\ref{fig:more15_gamma}, we find that the \citet{more_15} formula roughly matches the 75th percentile definitions at high peak heights.}
\label{fig:more15_nu}
\end{figure}

\citet{more_15} also provided formulae for the $\nu$--$\rsp$ and $\nu$--$\msp$ relations, again based on stacked density profiles. The dependence on $\nu$ was presumed to be due purely to the mass dependence of $\Gamma$. We compare their function to our results in Figure~\ref{fig:more15_nu}. As the \citet{more_15} function does not depend on redshift, it cannot match the trends found in Section~\ref{sec:results:nogamma}, but it once again coincides more or less with the 75th percentile $\rsp$ at high peak height. Interestingly, Figure 4 of \citet{more_15} shows a hint of the reversed redshift evolution (lowered \nursp at the highest redshifts), but the trend was not significant enough to be captured in their fitting function.

We conclude that there is no exact one-to-one match between the radius of the steepest slope and the definitions used in this paper. However, $\rspsf$ gives a good approximation, especially at high peak heights, where the results of \citet{more_15} were most constrained.

\subsection{Comparison with \shellfish}
\label{sec:comp:shellfish}

\begin{figure}
\centering
\includegraphics[trim = 2mm 6mm 3mm 0mm, clip, scale=0.62]{\figdir/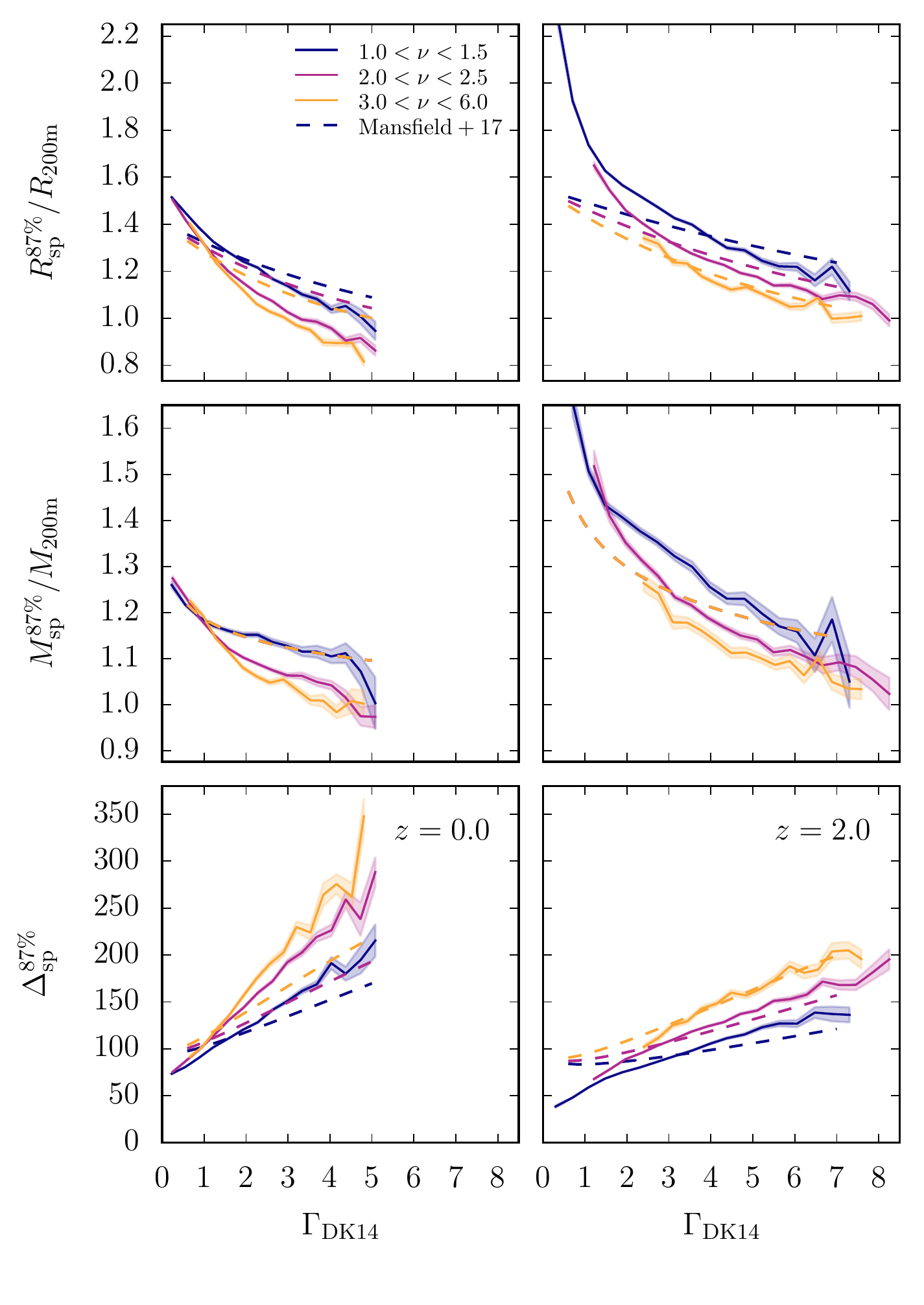}
\caption{Comparison of the median $\rsptom$ (top), $\msptom$ (middle), and $\deltasp$ (bottom) with the fitting function of \citet[][dashed lines]{mansfield_17}, where $\rsp$ and $\msp$ are defined as the $87$th percentile of the apocenter distribution. The two columns show redshifts of $0$ (left) and $2$ (right). The \shellfish relations are shown only for the range of $\Gamma$ where they were constrained. The fit to $\msptom$ does not depend on $\nu$, and the fit to $\deltasp$ is derived from the fits to $\rsptom$ and $\msptom$.}
\label{fig:mansfield}
\end{figure}
 
In \citetalias{diemer_17_sparta}, we undertook a halo-by-halo comparison of the results of \sparta and \shellfish. We found that $\rspes$ agrees with \shellfish to a few percent, though with about 15\% scatter. The small overall difference, however, does not mean that there could not be systematic trends with mass or accretion rate. Thus, we compare the \gammarsp measured by \sparta and \shellfish in Figure~\ref{fig:mansfield}. The relations agree reasonably well, but, driven by systematic differences at low $\Gamma$, \shellfish prefers values of $\rsptom$ that do not rise as sharply at low $\Gamma$. We further discuss the physical connection between splashback shells and apocenter distributions in Section~\ref{sec:discussion:physical}.

\subsection{Comparison with Theoretical Models}
\label{sec:comp:models}

\begin{figure}
\centering
\includegraphics[trim = 3mm 6mm 3mm 0mm, clip, scale=0.62]{\figdir/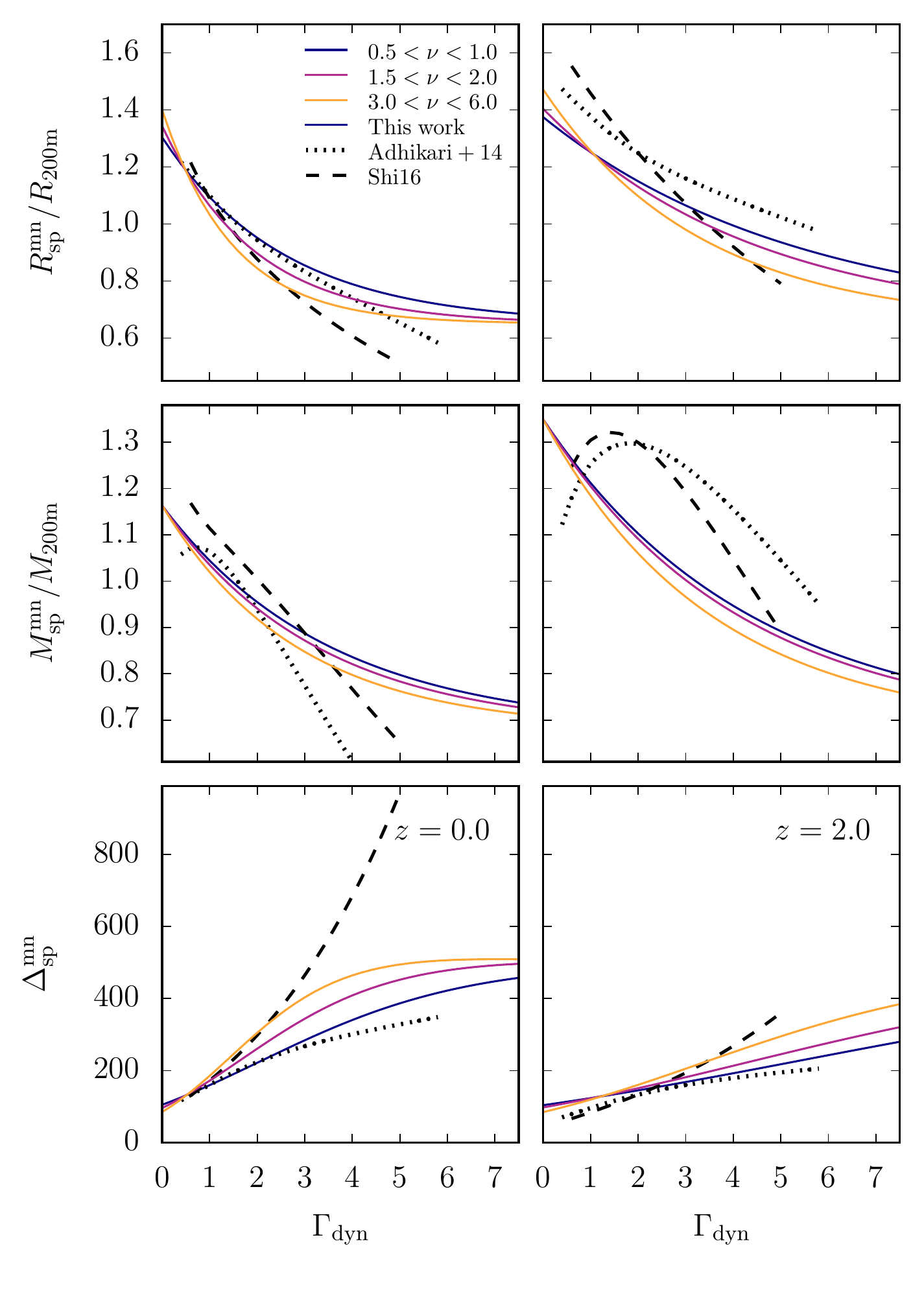}
\caption{Same as Figure~\ref{fig:mansfield}, but comparing the fitting function presented in this paper (solid lines, shown for a range of masses) with the analytical models of \citet[][dotted lines]{adhikari_14} and \citet[][dashed lines]{shi_16}. We are using $\rspmean$ for this comparison. In the analytical models, the mass accretion rate is understood to be instantaneous rather than measured over a dynamical time, which likely accounts for part of the differences. The models do not predict any dependence on halo mass.}
\label{fig:adhikari_shi}
\end{figure}

Given that the spherical collapse model provides the theoretical foundation for the splashback radius as a halo boundary, it is natural to use this type of model to predict $\rsp$ and $\msp$. In Figure~\ref{fig:adhikari_shi}, we compare the models of \citet{adhikari_14} and \citet{shi_16} to our results. Both models assume spherical symmetry, meaning that $\rsp$ is uniquely defined and that all particles reach their first apocenter exactly at $\rsp$. The values of $\rrsp$ measured by \sparta represent a distribution around this radius that is scattered due to the complexities of realistic structure formation. Assuming that this scatter does not bias the mean of the distribution, we use $\rspmean$ as the definition for this comparison. 

We caution that $\gammadyn$ is not equivalent to the instantaneous mass accretion rate $s$ used in the models, which assume $s = \mathrm{const}$ such that $M \propto a^s$. In this case, $\Gamma$ is equal to $s$ regardless of what interval $\Gamma$ is averaged over, but realistic halo mass accretion rates tend to decrease with time at low redshifts, meaning that an average such as $\Gamma$ likely overestimates the instantaneous accretion rate. Thus, the different definitions complicate the interpretation of the comparison shown in Figure~\ref{fig:adhikari_shi}.

\citet{adhikari_14} used a spherical shell collapse model \citep[e.g.,][]{gunn_72_sphericalcollapse}, assuming an NFW profile for the mass inside a given radius. The concentration is set by matching the slope of the profile at $\rvir = 1/2\ r_{\rm turn-around}$ to the spherical infall prediction, resulting in concentrations that do not necessarily match those observed in cosmological simulations. Due to this definition of the concentration, the prediction cannot be extended past $s = 6$. Given the NFW mass profile, \citet{adhikari_14} numerically computed the radius at which shells reach their first apocenter. The dotted lines in Figure~\ref{fig:adhikari_shi} show this prediction.\footnote{We do not show the fitting function given in \citet{adhikari_14}, but rather the results of an improved numerical calculation of $\deltasp$ (S. Adhikari, private communication). We compute $\rsptom$ and $\msptom$ from an NFW profile with the concentration used in the model. We have implemented both this method and the model of \citet{shi_16} in the Python module \colossus.}

Instead of assuming a particular function for the density profile, \citet{shi_16} performed a self-consistent calculation of shell collapse. As in the predecessor models of \citet{fillmore_84} and \citet{bertschinger_85}, this calculation results in a power-law density profile with a sharp drop-off at the splashback radius. The power-law slope depends on the slope of the initial perturbation, which also sets the accretion rate. The model takes $\Omega_{\rm m} \neq 1$ into account and thus makes redshift-dependent predictions. However, due to the self-similarity of the problem at fixed redshift, the model predicts no mass dependence. The black dashed lines in Figure~\ref{fig:adhikari_shi} show fits to the numerical model predictions that were given for $\rsptom$ and $\deltasp$, and reconstructed for $\msptom$. We note that the downturn in $\msptom$ at low $\Gamma$ is present in the model but exaggerated due to the reconstruction from the fits to $\rsptom$ and $\deltasp$ (X. Shi, private communication). The fitting functions were constrained for mass accretion rates in the range $0.5 < s < 5$ \citep{shi_16}.

Both models correctly predict the general trend of a decreasing $\rsptom$ with mass accretion rate. In detail, however, the models do not match our results: the slope of the \gammarsp is steeper at low redshift and the normalization is higher at high redshift. The evolution of $\msptom$ with $\Gamma$ is also steeper at low redshift and entirely different at high redshift. We discuss the physical reasons for these disagreements in Section~\ref{sec:discussion:theoretical}.
 
\section{Discussion}
\label{sec:discussion}

We have analyzed our results for the splashback radius, mass, and overdensity as a function of halo mass, accretion rate, redshift, and cosmology and expressed them in a convenient fitting function. In this section, we physically interpret some of the results and discuss the theoretical implications.

\subsection{On the Physical Meaning of the Apocenter Distribution}
\label{sec:discussion:physical}

The \sparta algorithm provides a new, independent way to measure the splashback radius, adding to two previous definitions (the radius where the density profile is steepest and the nonspherical splashback shell determined by \shellfish). In the spherical collapse model, all of these definitions are equivalent, because a spherical shell reaches apocenter at a fixed time, and all particles splash back at the same time and radius. This pileup causes the sharp density drop in the model \citep{fillmore_84, bertschinger_85}. 

In reality, the situation is more complicated. First, nonsphericity causes an intrinsic scatter in the apocenter radii \citep{adhikari_14}. Second, the energy and angular momentum of particles at infall slightly influences their apocenter \citepalias{diemer_17_sparta}. One could argue that these effects can be seen as adding scatter but not shifting the mean apocenter, and that $\rspmean$ should thus be the best definition of $\rsp$. However, it is not a priori clear that $\rspmean$ has any signatures that are observable in the real universe, meaning we need to establish a connection to the properties of the density profile.

The density profile does not carry a unique signature of the ``true'' splashback radius either. While the inner profile falls steeply near the splashback radius, the density due to nonlinearly infalling shells becomes increasingly important with radius\footnote{We note that this density contribution is not well described by the so-called 2-halo term, i.e. the statistical contribution due to the clustering of halos \citep[e.g.,][]{smith_03_powerspec, hayashi_08, tinker_10_bias}. This contribution begins to dominate only at much larger radii \citep{diemer_14_profiles}.}. Thus, the location of the steepest density slope represents a trade-off between the inner and outer profiles and cannot trivially be interpreted as the splashback radius. In observations, however, this radius is the most accessible quantity, and we have shown that it is reasonably approximated by $\rspsf$ as measured by \sparta. This connection will be investigated in more detail in future work, ideally using hydrodynamical simulations to directly connect the apocenter distribution to observables such as the density of satellite galaxies in clusters.

The fact that the radius of the steepest slope is merely one possible definition of $\rsp$ was illustrated by the results of \citet{mansfield_17}, who found that massive substructures bias the radius of the steepest slope by about 30\% compared to measurements where substructure has been removed. As a result, they find a somewhat larger $\rsp$ on average, even though their measurements are also based on the density field. Another effect contributing to this difference may be the nonspherical nature of their shells (which is converted to a volume-equivalent radius). We have identified $\rspes$ as the best proxy for the \shellfish results \citepalias{diemer_17_sparta}, but the two measurements differ significantly in some halos, which remains to be investigated in more detail. 

In summary, there is no one definition of $\rsp$ that clearly corresponds to the density drop in the spherical collapse model and that can be measured in both simulations and observations. In the future, we hope to establish a tighter connection between the density drop measured by \shellfish and the apocenter distribution by considering the three-dimensional distribution of apocenters rather than only their radii.

\subsection{On the Relationship between Splashback and Spherical Overdensity}
\label{sec:discussion:universality}

Perhaps the most striking feature of the data in presented in this paper, and thus the fitting function given in Equations~(\ref{eq:fit1})--(\ref{eq:fit3}), is their relative complexity, with significant dependencies on halo mass and $\Omega_{\rm m}$ (at fixed $\Gamma$). While the latter dependence was expected from theoretical considerations \citep{adhikari_14, shi_16}, the halo mass dependence was not (though it was recently found in simulations by \citealt{mansfield_17}). These complexities raise the question of whether we expect there to be a simple relation between $\rsp$ and conventional spherical overdensity radii. We note that $\rsp/\rdelta$ would vary more strongly if spherical overdensity radii other than $\rtom$ were used.

At a fixed $\rsp$, the ratio $\rsptom$ depends on the mass profile around $\rsp$ which depends on mass, accretion rate, and redshift in a nontrivial way \citep[e.g., Figures 3 and 10 in][]{diemer_14_profiles}. These dependencies are expected from the fact that concentration depends on mass, redshift, and cosmology in a complex fashion \citep[e.g.,][]{bullock_01_profiles}. We expect a significant correlation between concentration and mass accretion rate \citep{wechsler_02_halo_assembly}, raising the question of whether $c$ could be substituted for $\Gamma$ in our fitting model.
 
Another open question is related to the dependence of $\rsptom$ on cosmological parameters. We have shown that the scaling with $\Omega_{\rm m}$ works for the $WMAP$ and $Planck$ cosmologies, and is thus likely appropriate for any realistic cosmology. However, $\rsptom$ varies significantly between self-similar simulations that are distinguished only by their power-spectrum slope $n$, meaning that the splashback radius is sensitive to cosmological parameters beyond $\Omega_{\rm m}$. An alternative way to frame such issues could be to ask whether we are considering the optimal variables. For example, we quantify the mass accretion rate as $\gammadyn$, but perhaps $\rsp$ exhibits tighter correlations with other definitions that we have not yet considered (e.g. definitions based on shorter or longer timescales, or definitions relying directly on $\msp$ instead of $\mtom$). We will systematically explore the correlations with other parameters (such as concentration) in future work.

On a theoretical level, one of the most important differences between $\rsp$ and conventional definitions is the meaning of the overdensity $\Delta$. In the context of spherical overdensity radii, $\Delta$ is the fundamental quantity that determines $\rdelta$ and $\mdelta$. In the splashback picture, $\deltasp$ is merely a consequence of independently determined $\rsp$ and $\msp$. Our results show that not only does $\deltasp$ vary systematically depending on a number of variables but also that the scatter in $\deltasp$ is larger than the scatter in $\rsp$ and $\msp$, highlighting that there is nothing fundamental about the splashback overdensity. This difference in interpretation has a bearing on the physical interpretation of halo growth. For example, due to the constant $\Delta$, $\rdelta$ can change suddenly when a massive subhalo is accreted. In contrast, $\rsp$ changes more slowly in this case, while $\msp$ and thus $\deltasp$ change rapidly.

In summary, we have little reason to expect a simple relation between $\rsp$ and $\rdelta$. While we employ the quantities $\rsptom$ and $\msptom$ to establish a connection between $\rsp$ and $\rdelta$, $\rsp$ is an independent definition of the halo boundary that cannot easily be expressed in terms of spherical overdensity radii.

\subsection{Compatibility with Observations}
\label{sec:discussion:obs}

The measurement of $\rsp$ performed by \citet{more_16} and confirmed by \citet{baxter_17} indicates a surprisingly small splashback radius, namely, $\rsptom = 0.837 \pm 0.031$ for a cluster sample with $\nu = 2.4$ at $z = 0.24$. While we cannot directly measure the mass accretion rate of the clusters, Figure~\ref{fig:fits} clearly shows that the theoretically expected value is higher. 

In particular, the fitting function of \citet{more_15} predicts $\rsptom = 1.1$ for this sample, 32\% higher than the observed value. The fitting function presented in this paper predicts almost exactly the same value for $\rspsf$ at the given peak height and redshift. Given the scatter in the \nursp, the observed $\rsp$ would represent a 2$\sigma$ fluctuation even for an individual halo, whereas the \citet{more_16} result was derived by stacking the density profiles of thousands of clusters. 

In contrast, \citet{busch_17} showed that the observed location of the steepest slope of the density profile can be sensitive to the details of the cluster identification algorithm. In addition, the assumption of spherical symmetry can affect the inferred radius of the steepest slope in three dimensions. Thus, the significance of the disagreement between simulations and observational results will remain unclear until the cluster-finding algorithm can be tested with realistic mock catalogs.

\subsection{The Status of Analytical Models}
\label{sec:discussion:theoretical}

In Section~\ref{sec:comp:models}, we found that none of the semi-analytical models that have been proposed to date predict our results in detail. The model of \citet{shi_16} corresponds to the prediction of the spherical collapse model in a $\Lambda$CDM universe. Due to the self-similarity of the setup, this model predicts a power-law inner density profile, $\rho \propto r^\alpha$, whereas the density profiles in cosmological simulations steepen with radius. Given that the \citet{adhikari_14} model is based on an NFW profile instead of a power law, the relative similarity of the predictions for $\msptom$ might seem surprising. The agreement can be explained by the \citet{adhikari_14} procedure for setting the NFW concentration: the slope of the mass profile is set to the spherical collapse model prediction of $3s/(3+s)$, even at $s > 3/2$, leading to density slopes of $\alpha \to -3$ when $s \to 0$ (and thus $c \to \infty$) and $\alpha \to -1$ when $s \to 6$ (and thus $c \to 0$). These extreme values of concentration mean that the NFW profile in the \citet{adhikari_14} model approaches a power-law shape for both low and high accretion rates.

In summary, spherically symmetric models of shell collapse are a promising class of models for predicting the splashback radius. However, these models need to be coupled with realistic halo density profiles in order to match the simulation results in detail. One possible avenue toward more accurate models would be to set the concentration of the inner profile according to a numerically calibrated concentration--mass relation, automatically introducing a dependence on halo mass that is not present in self-similar collapse models.


\section{Conclusions}
\label{sec:conclusion}

Using the \sparta algorithm described in \citetalias{diemer_17_sparta}, we have quantified the splashback radii and masses of a large sample of halos from $N$-body simulations of different $\Lambda$CDM cosmologies. We have investigated the dependence of those quantities on halo mass (expressed as peak height, $\nu$), accretion rate $\Gamma$, redshift (expressed as $\Omega_{\rm m}$), and cosmology. The relatively complex dependencies indicate that the splashback radius represents an independent definition of the halo boundary that cannot simply be reconstructed from conventional spherical overdensity definitions. Our main conclusions are as follows.
\begin{enumerate}
\item At fixed $\Gamma$, mass, and redshift, $\rsp$ and $\msp$ are distributed roughly log-normally, with some tails toward high values. The 68\% scatter in both $\rsp$ and $\msp$ varies between about $0.02$ and $0.1$ dex, where scatter decreases with $\nu$ and increases with $\Gamma$. If the accretion rate is unknown and we average over all $\Gamma$, the distribution is still close to log-normal, but the scatter in the relations increases to about $0.07$ dex in $\rsp$ and between $0.04$ and $0.1$ dex in $\msp$.

\item In agreement with previous work, we find that $\rsptom$ and $\msptom$ decrease with accretion rate and increase with $\Omega_{\rm m}$, but we also find a significant dependence on $\nu$. We do not find any dependence on cosmological parameters (beyond the dependence on $\Omega_{\rm m}$) within different $\Lambda$CDM cosmologies. We parameterize the median $\rsp$ and $\msp$ as a function of $\Gamma$, $\nu$, and $\Omega_{\rm m}$, which is accurate to 5\% or better for all masses $\mtom > 1.7 \times 10^7 \msunh$, up to $z = 8$, and for the $WMAP$ and $Planck$ cosmologies that span the currently favored range of cosmological parameters. This function is implemented in the publicly available Python code \colossus.

\item We give a fitting function for the accretion rate as a function of $\nu$ and $z$. Using the fitted $\Gamma$ as input to the fitting functions for $\rsp$ and $\msp$, we obtain predictions for the $\nu$--$\rsp$ and $\nu$--$\msp$ relations that are accurate to 5\% or better.

\item We compare our results to measurements of $\rsp$ from stacked halo density profiles and find that the radius of the steepest slope as measured by \citet{more_15} corresponds roughly to the 75th percentile of the splashback distribution of particles. Similarly, we find that the 87th percentile results in a \gammarsp that is roughly compatible with that measured using the \shellfish code of \citet{mansfield_17}.

\item We compare our results to several semi-analytical models of $\rsp$. While these models reproduce the general trends of $\rsp$ and $\deltasp$, they do not predict any mass dependence and cannot explain our results in detail. 
\end{enumerate}
Rather than a definitive statement, the results presented in this paper represent a starting point in our investigation of the splashback radius as a viable alternative to conventional radius definitions. Our goal is to create self-consistent halo catalogs with $\rsp$ measurements or estimates for all halos above a certain mass threshold and subhalo relations based on $\rsp$. With such catalogs, a number of classical topics in structure formation can be re-visited, namely, semi-analytical models of galaxy formation, assembly bias, or the growth of halos and its connection to concentration. Furthermore, our theoretical understanding of $\rsp$ and $\msp$ is still lacking, as an accurate analytical description of our results from first principles remains elusive.


\vspace{0.5cm}

We are grateful to Susmita Adhikari, Neal Dalal, and Xun Shi for enlightening discussions. BD gratefully acknowledges the financial support of an Institute for Theory and Computation Fellowship. This work was completed using the \textsc{Midway} computing cluster provided by the University of Chicago Research Computing Center. 


\bibliographystyle{aasjournal}
\iflocal
\bibliography{../sf.bib}
\else
\bibliography{sf.bib}
\fi

\end{document}